\documentclass{ametsocV6.1}
\usepackage{lineno}
\AtBeginDocument{\nolinenumbers}
\nolinenumbers
\makeatletter

\renewcommand\internallinenumbers{}
\renewcommand\resetlinenumber[1]{}

\makeatother
\usepackage{url}
\usepackage{siunitx}%
\usepackage{cleveref}

\title{Rare events algorithm study of extreme double jet summers and their connection with heatwaves over the Northern Hemisphere}

\authors{Valeria Mascolo,\aff{a}\correspondingauthor{Valeria Mascolo, v.mascolo@reading.ac.uk}
Francesco Ragone,\aff{b}
Nili Harnik,\aff{c}
Freddy Bouchet,\aff{d}
}

\affiliation{\aff{a}{Laboratoire de Physique à l'ENS de Lyon, CNRS, F-69342 Lyon, France}\\
\aff{b}{School of Computing and Mathematical Sciences, University of Leicester, LE17RH Leicester, United Kingdom}\\
\aff{c}{Department of Geophysics, Tel Aviv University, Tel Aviv, Israel}\\
\aff{d}{Laboratoire de Météorologie Dynamique, IPSL, ENS-PSL, CNRS, Paris, France}
}

\abstract{Several large scale circulation patterns have been identified in relation to extreme Northern Hemisphere summer heatwaves. Three main ones are a double jet over Eurasia, a positive phase of the summer northern annular mode, and a quasi-wave-3 geopotential height anomaly. While there is some evidence suggesting these patterns are related to each other, the explicit nature of their relation, as well as the explicit mechanisms by which they are related to extreme heatwaves is still not known. The double jet structure has gained attention recently due to evidence that its persistence has been increasing, possibly explaining the rise in the number of extreme heatwaves over Europe. In this paper we study the occurrence and persistence of double jet states in ERA5 and in stationary simulations with the CESM1.2 model, using an index which measures the degree of jet separation. Additionally, we perform simulations with CESM1.2 coupled to a rare event algorithm in order to improve the statistics of rare summer-long double jet states. We find that extreme double jet states are characterised by three centers of extreme high surface temperature and 500hPa geopotential height anomalies, alongside a strong low pressure over the Arctic. The geopotential height anomaly pattern is consistent with both a positive Northern Annular Mode (NAM) and quasi-wave-3 patterns found in the literature. Moreover, we find a large percentage of co-occurrence of heatwaves at these centers, and a double jet state, with the percentage increasing with the duration of the double jet state. }

\begin{document}

\maketitle

\section{Introduction}

Long-lasting midlatitude heatwaves are among the extreme events with the largest impacts on society, and there is high confidence that their frequency and amplitude will increase in response to climate change \citep{intergovernmental_panel_on_climate_change_climate_2023}.
Several mechanisms have been identified in the formation of midlatitude  heatwaves~\citep{horton_review_2016,barriopedro_heat_2023}, including land-atmosphere feedbacks \citep{miralles_landatmospheric_2019,miralles_mega-heatwave_2014,seneviratne_landatmosphere_2006}, sea surface temperature anomalies \citep{della-marta_summer_2007,cassou_tropical_2005}, and large-scale circulation patterns \citep{cassou_tropical_2005,della-marta_summer_2007}.  

Focusing on Northern Hemisphere summer, several global-scale patterns have been associated with extreme heatwaves, especially over Europe. \citep{ogi_summer_2005} related the extreme hot summer of 2003 to a strongly positive summer Northern Annular Mode (NAM). \citep{ragone_rare_2021} found that extreme persistent heatwaves over France and Scandinavia are associated with a large scale high pressure system over that region, which is part of a planetary scale quasi-wave-3 teleconnection pattern. \citep{rousi_accelerated_2022} found that an extremely persistent  Eurasian double jet pattern favourable for the formation of extreme heatwaves. 

Understanding the large scale patterns which enhance the occurrence of extreme heatwaves, on a range of persistence scales, is central to understanding how global warming might affect their occurrence beyond a simple shift of the distribution to a warmer mean. Specifically, it is not clear how the three global scale patterns above are related, but there is reason to think that they are. \citep{ogi_summer_2005} and \citep{tachibana_abrupt_2010} noted that the positive phase of the summer NAM is characterized by a split jet over the Atlantic-Eurasian region. An examination of the \SI{500}{\hecto\pascal} geopotential height anomaly during the positive summer NAM~\citep{ogi_summertime_2004} shows similar Western Europe and North America persistent large high pressure anomalies as in the quasi-wave-3 pattern of~\citep{ragone_rare_2021}. In addition, the upper tropospheric meridional wind anomalies during the most persistent double jet summers~\citep{rousi_accelerated_2022} suggests the presence of an anomalous high latitude quasi-wave-3 pattern. 

Several dynamical links between the large scale circulation patterns and heatwaves have been suggested. 
The positive phase of the summer NAM was shown by \citep{tachibana_abrupt_2010} to be associated with significantly more blocking highs, which are associated with warm extremes~\citep{pfhal_wernli_2012}. A different mechanism is through an enhancement of medium-scale Rossby wave trains, which have been associated with the formation of extreme heatwaves in different parts of the world~\citep{hoskins_persistent_2015,kornhuber_extreme_2019,teng_probability_2013,mann_influence_2017}, including Europe \citep{coumou_quasi-resonant_2014,kornhuber_evidence_2017,kornhuber_summertime_2017}.  \citep{rousi_accelerated_2022} suggested that the recent increase in the occurrence of extreme European heatwaves is linked to an increase in the persistence of double jet states. 

The explicit mechanisms by which a double jet state is related to extreme heatwaves is not well understood. 
On the one hand, the polar jet can be forced by the strengthening of the land-sea temperature gradient along the Arctic sea coast~\citep{coumou_influence_2018}, suggesting the increase in extreme heatwaves results from the tendency of a second high latitude jet peak to form. On the other hand, the splitting of the jet can arise from the formation of large flow undulations, as is the case in a strong blocking high~\citep{wirth_problem_2021}, meaning the co-occurence with heatwaves could be incidental rather than causal. 

The double jet structure is inherently linked to the summer NAM on synoptic to subseasonal time scales. \citep{tachibana_abrupt_2010} examined the life cycle of abrupt onsets of a positive summer NAM, and showed how wave-mean flow interactions lead to a splitting of the jet during such episodes, as well as the preferred development of blocking highs.  
While the summer NAM related wave-mean flow feedbacks gave rise to double jet events on a time scale of 10 days or so \citep{tachibana_abrupt_2010}, externally forced double jets \citep{coumou_influence_2018}, as well as land-surface feedbacks can excite more persistent double jet states \citep{miralles_mega-heatwave_2014}. 

Thus, understanding the dynamics of the double jet structure seems important to understanding future NH summer climate change and extreme event distribution. 
In this paper we focus on the double jet structure, and specifically on persistent summer-long double jets. We analyse ERA5 data \citep{hersbach_era5_2020} to address the following questions:

$\bullet$ How often have seasonal double jet states occurred, and how pronounced is the splitting during these summers?

$\bullet$ Does a persistent double jet favor long-lasting heatwaves? 

$\bullet$ Is this persistent state associated with persistent summer NAM and quasi-wave-3 patterns?

A key difficulty in answering the above questions from observations is that the historical records are in general too short to observe a significant sample of such events. Similarly, studying these events with numerical simulations is computationally extremely demanding.  

The issue of computational costs can be tackled using rare event algorithms. These are computational techniques whose goal is to oversample rare events of interest in ensemble simulations, allowing to gather sufficient statistical data at a lower computational cost than direct simulations \citep{del_moral_genealogical_2005,rolland_computing_2016}. Rare event algorithms have been recently applied in climate science to study warm summers over France and Scandinavia \citep{ragone_computation_2018,ragone_computation_2020,ragone_rare_2021}, and India \citep{le_priol_using_2024}, mid-latitudes precipitations \citep{wouters_rare_2023}, weakening and collapse of the Atlantic Meridional Overturning Circulation \citep{cini_simulating_2024}, melting of the Arctic sea ice \citep{sauer_extremes_2024} and for energy demand in the power system \citep{cozian_computing_2023}. 

In addition to the analysis on ERA5 data, in this study we perform experiments with a rare event algorithm applied to the CESM1.2 climate model \citep{hurrell_community_2013} to sample double jet structures over the Eurasian continent. The algorithm is implemented in order to select model trajectories characterized by high values of an index able to identify double jet structures. The application of the algorithm allows to simulate double jet structures with return times orders of magnitude larger than what is feasible with direct sampling, and to calculate statistically significant composite maps of physical quantities conditional on the occurrence of these dynamical structures. 

The paper is organised as follows. In \cref{sec:dataandmethod} we present the data and models used for this study and we define the double jet index used to identify the atmospheric double jet structures. We then introduce the rare event algorithm and discuss the setup of the experiments. In \cref{sec:results_dji} we first show the performance of the double jet index on a control run of CESM1.2.2 and on the reanalysis dataset ERA5. Then, we show the performances of the the rare events algorithm in sampling rare persistent double jet states, and we analyse their connection to heatwaves. In \cref{sec:conclusions_dji} we summarise our findings and discuss future perspectives.

\section{Data and methods}
\label{sec:dataandmethod}

\subsection{Data}

 
    \subsubsection{ERA5} 
    \label{subsec:era5}

    We use  the public available reanalysis dataset ERA5 \citep{hersbach_era5_2020}. Specifically, we use daily data for the summer months of June, July and August, starting from 1940 until 2022 for the \SI{2}{\meter} air temperature ($T_\text{2m}$), \SI{500}{\hecto\pascal} geopotential height ($Z_\text{500}$), zonal wind $U$ (averaged between \SI{200}{\hecto\pascal} and \SI{350}{\hecto\pascal}). The dataset has a resolution of 0.25 degree both for latitude and for longitude. In order to remove the effect of global warming, we performed a latitudinal-wise detrending of both \SI{2}{\meter} air temperature, \SI{500}{\hecto\pascal} geopotential height longitudinally averaged over the Northern Hemisphere. 
    The procedure is detailed in 
    \cref{supmat:era}.

    \subsubsection{CESM1.2}
    \label{subsec:cesm}

    We use CESM version 1.2.2 \citep{hurrell_community_2013} in an atmosphere-land only setup, with prescribed sea surface temperatures, sea ice distribution and fixed greenhouse gases concentrations ($CO_2$ concentration set at 367 ppmv) to match the average climate conditions of the 2000s. The atmospheric model is the Community Atmospheric Model version 4 (CAM4), while the Community Land Model version 4 (CLM4) is used for the land. The model has a resolution of 0.9 degrees in latitude and 1.25 in longitude, with 26 vertical layers in a hybrid pressure-sigma coordinate. 
    We analyze  a control simulation 1000 years long, which has been already used to study heatwaves and energy production in Europe \citep{ragone_rare_2021, cozian_assessing_2024}, and we perform a set of experiments  coupling CESM1.2 to a rare event algorithm to sample rare double jet events. 
    The procedure is presented in \cref{subsec:rea} and the results in \cref{sec:rea_results} and \cref{sec:rea_teleconection}. The analysis is based on daily values of zonal wind $U$ between 192 and 313 hybrid sigma pressure levels, \SI{2}{\meter} temperature ($T_\text{2m}$) and \SI{500}{\hecto\pascal} geopotential height ($Z_\text{500}$).

\subsection{Methods}
    \subsubsection{Double jet index}
    \label{subsec:dji_def}

    We identify double jet structures in the Northern Hemisphere using a variation of an index developed for the Southern Hemisphere in  
    \citep{yang_variability_2006}, in which we take the difference between the zonal wind anomalies, averaged over a box taken around the poleward peak of the double jet and a box taken around the inter-jet region. The zonal mean zonal wind variability is such that the value of this index is maximal when the jet is split, and minimal when there is a single jet. To best capture the jet core, we calculate this index from the daily zonal wind $U$ anomaly (with respect to the daily, grid point-wise climatology of the months of June, July and August) vertically averaged over the 4 upper-tropospheric levels which lie between 192 and 313 hPa (pressure levels in ERA5 and sigma levels in CESM1.2).

\begin{figure}
        \centering
        \includegraphics[width=0.8\textwidth]{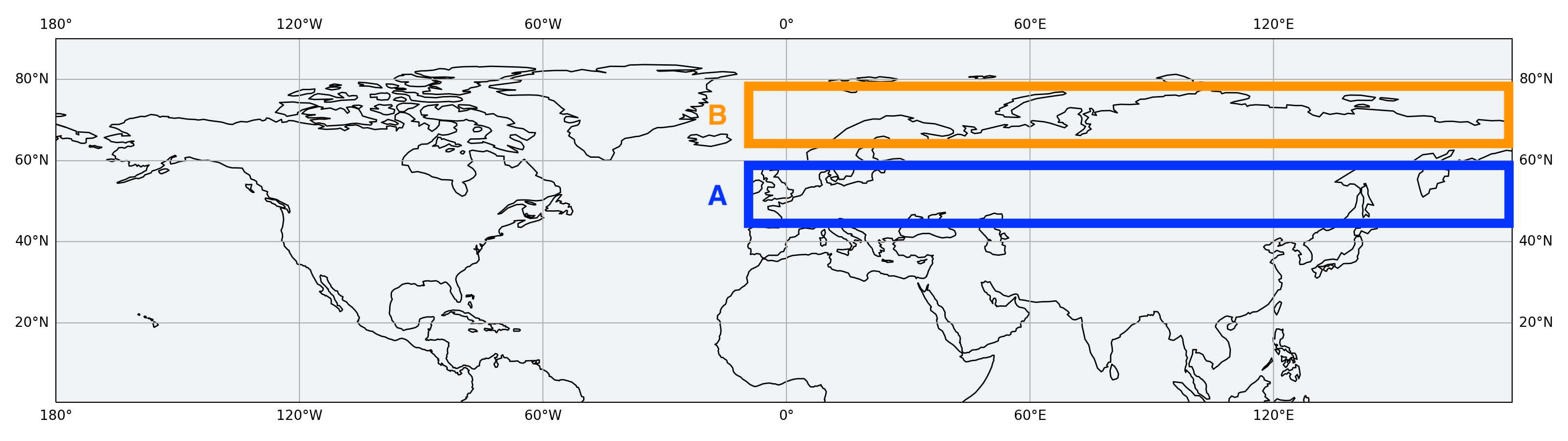}
        \caption{Example of the two zones used to define a double jet: $B$ between the latitudinal band 65-78.5$^\circ$N and $A$ between 46.5-60$^\circ$N.}
        \label{fig:zones-dji}
    \end{figure}

     To also allow for small meridional variations in the latitude of the poleward peak and inter-jet region, we calculate the maximal index for a range of possible central latitudes for each of the boxes, as follows. 
    We stress that this is a novelty we introduced given the latitudinal dependency of the poleward peak of the jet (not shown).
    We denote the inter-jet zone by $A$, and the poleward peak zone as $B$. We define the latitudinal width of the boxes to be $w = 13.5^\circ$. Box $A$ lies somewhere between  40$^\circ$N and 65$^\circ$N, while box $B$ lies somewhere between 60$^\circ$N and 85$^\circ$N. We then take a zonal average over the longitudinal sector corresponding to Eurasia, from 10$^\circ$W to 180$^\circ$E and calculate all the possible box $B$ minus box $A$ pairs for box-latitudes of difference $d=5^\circ$ apart, and find the maximal value. Thus, we define then our double jet index $D$ as the daily maximum of this difference:
    \begin{equation}
        D(t) := \operatorname*{max}_{\tilde{\lambda} \in (40,52)} \left[ \int_{\tilde{\lambda} + w + d}^{\tilde{\lambda} +2w + d}U(t,\lambda)d\lambda - \int_{\tilde{\lambda} }^{\tilde{\lambda} + w}U(t,\lambda)d\lambda\right ]
    \end{equation}
    where $U(t,\lambda)$ is the zonal wind daily anomaly already averaged over the vertical levels and the Eurasian longitudinal sector (from 10$^\circ$W to 180$^\circ$E), $t$ denotes the time and $\lambda$ the latitude. The first integral corresponds to zone $B$, while the second to zone $A$. 
    For illustration, \cref{fig:zones-dji} 
    shows an example of the two zones for an optimal combination of box $B$ between 65-78.5$^\circ$N and box $A$ between 46.5-60$^\circ$N. To take into account persistence, we perform a moving window average of the daily double jet index with fixed-durations windows, ranging from 7 up to 90 days, as shown later on in the manuscript. 

    The resulting index is not sensitive to the order in which we take the vertical, zonal and latitudinal averages. 
    
    The results are also not sensitive to reasonable variations of the vertical averaging levels and geographical averaging areas, and whether we use wind speed instead of the zonal wind component only (not shown). We note that other definitions have been proposed to detect this splitting \citep{molnos_network-based_2017, rousi_accelerated_2022, gallego_new_2005, pena-ortiz_observed_2013}. 
    The version presented here is efficient in identifying double jet structures, while remaining computationally inexpensive to calculate for a given atmospheric configuration. 

    \subsubsection{Heatwave definition}
    \label{subsec:hw_def}
    In this work we are interested in heatwave events which are persistent in time and space. Let $\tilde{T}_\text{2m}$ denote the daily-averaged \SI{2}{\meter} air temperature field, which depends on the location $\vec{r}$ and time $t$. We consider the temperature anomaly $T_\text{2m} := \tilde{T}_\text{2m} - \mathbb{E}_y(\tilde{T}_\text{2m})$ where $\mathbb{E}_y(\tilde{T}_\text{2m})$ is the local climatology. We consider the space and time average $A$ of the temperature anomaly
    \begin{equation} \label{eq:heatwave}
        A(t) := \frac{1}{T}\int_{t}^{t+T} \left( \frac{1}{\mathcal{A}}\int_{\mathcal{A}}T_\text{2m}(\vec{r},u)d\vec{r}  \right)du,
    \end{equation}
    where $T$ is the duration and $\mathcal{A}$ is the spatial region of interest. We define a heatwave event of length $T$ over $\mathcal{A}$ for values of $A$ above the 95th percentile. 

    This definition of heatwave events has been used in various studies \citep{galfi_large_2019, galfi_fingerprinting_2021, ragone_computation_2018, ragone_rare_2021, jacques-dumas_deep_2022, miloshevich_robust_2023}, and has the advantage of depending explicitly on the intensity and duration of the event, which are then parameters of the analysis. In the following we will vary $T$ from sub-weekly to monthly time scales, enabling to study short and long events. We will consider areas with spatial extension comparable to the synoptic scales, which is of the order of \SI{1000}{km} at the midlatitudes, as these scales correlate to the size of cyclones and anticyclones and jet stream meanders. In particular, we will focus on 3 regions in the Northern Hemisphere, shown in 
    \cref{fig:hw_regions}. 
    The choice of these regions is motivated by teleconnection pattern discussed in \cref{sec:rea_teleconection}. The analysis will be limited to the summer months of June, July and August (JJA). 

    \subsubsection{The rare event algorithm}
    \label{subsec:rea}
    
    We use a rare event algorithm already employed to study heatwaves over France and Scandinavia \citep{ragone_computation_2018, ragone_rare_2021}, heatwaves in India \citep{le_priol_using_2024}, energy demand in the electric power system \citep{cozian_computing_2023}, the collapse of the Atlantic Meridional Overturning Circulation \citep{cini_simulating_2024} and Arctic sea ice reduction \citep{sauer_extremes_2024}, commonly known in the literature as GKLT (Giardinà-Kurchan-Lecomte-Tailleur) algorithm. 
    
    The goal of the algorithm is to sample more efficiently the tail of the distribution of a target observable $O[X(t)]$ in a numerical simulation. An ensemble of trajectories is run in parallel and at constant intervals of a resampling time $\tau$, a weight is assigned to each of them, based on the value of the target observable $O[X(t)]$ averaged over the previous time period $\tau$. 
    Based on this weight, trajectories achieving a low value are killed, while trajectories having a high score are cloned, repopulating the ensemble. A small perturbation is added to the clones before restarting the simulation for the next time $\tau$. The idea is that this selection process will favor the survival of trajectories leading to extreme events characterized by time persistence of large values of $O[X(t)]$, such as the double jet events we are interested in. More details can be found in \citep{ragone_computation_2018,ragone_rare_2021}.

   Let us consider a realisation of a climate simulation, denoted from now on as a trajectory $\{X(t)\}_{t_a \leq t \leq t_a +T} $, where $t_a$ is the initial starting point of the simulation and $T$ the total duration. We denote as $\mathbb{P}_0 \left( \{X(t)\} \right)$ the probability of observing a certain trajectory as a realisation of the dynamics of the climate model and as $\mathbb{P}_k \left( \{X(t)\} \right)$ the probability of observing the same trajectory as a result of the ensemble simulations driven by the rare event algorithm. Following  \citep{ragone_computation_2018, ragone_rare_2021}, there exists a link between the two probabilities:
    \begin{equation}
    \label{eq:pk_p0}
        \mathbb{P}_k \left( \{X(t)\} \right) = \frac{\exp{\left( k \int_{t_a}^{t_a + T} O[X(t)] dt \right)}}{Z} \mathbb{P}_0 \left( \{X(t)\} \right)
    \end{equation}
    
    where $Z$ is a normalization constant, $O[X(t)]$ is the target observable and $k$ the biasing parameter, a parameter which controls the selection strength, i.e. the higher its value, the larger values of the integrated time average of $O[X(t)]$ will be sampled. \Cref{eq:pk_p0} allows having access to the probabilities of the real model statistics (and to related quantities, such as averages) using the rare event algorithm simulated by inverting it. 
    
    The target variable for this study is the integrated double jet index averaged over Eurasia presented in \cref{subsec:dji_def},  $O[X(t)] \equiv D (t)$. We perform $M = 10$ ensemble simulations with the rare event algorithm, each of them including $N= 100$ trajectories starting from June 1st and ending on August 29th (duration $T = 90$ days), with selection strength $k = 0.01 (m s^{-1})^{-1}(day)^{-1}$ and resampling time $\tau = 5$ days.
    
    The 10 experiments start from 10 independent sets of 100 independent initial conditions, i.e. we took $1000$ independent June 1st, at 1-year interval. The chosen integration time allows us to analyse extreme double jets throughout the summer period.  The value of $k$ has been chosen such that events with a return time of 100 years are expected to become common, following a scaling argument presented in \citep{ragone_rare_2021}. This value has been chosen after some sensitivity tests and the value proposed here assures an average value of 10\% of the total number of trajectories in each experiment with independent ancestors at the beginning of the simulation (see \cref{supmat:ttest}). The resampling time $\tau$ in general should be of the order of the autocorrelation time of the target observable. An analysis of the autocorrelation function of the integrated double jet index (
    \cref{fig:acf-dji})
    shows that it can be approximated by a double exponential function, with a first decay time scale of $5$ days. The computational cost of the experiments with the algorithm is equivalent to simulating 1000 summers in the ensemble control run, but they allow to gather a much richer statistics for the extreme events of interest. 

\section{Results}
\label{sec:results_dji}
In \cref{subsec:dji_cesm_era} we show that our definition of double jet correctly represents this atmospheric state in both the CESM1.2 and ERA5 datasets. We analyze the relationship between double jets and heatwaves and double jets and persistent cyclonic anomalies in the polar region in \cref{sec:dji_hw,sec:dj_am}, respectively. In \cref{sec:rea_results,sec:rea_teleconection} we simulate extreme double jet summers using a rare events algorithm and present the results for teleconnection patterns and associated long-lasting heatwaves.

    \subsection{Characteristics of the double jet in ERA5 and CESM1.2} 
    \label{subsec:dji_cesm_era}
    
\begin{figure}[t]
        \centering
        \includegraphics[width=1\textwidth]{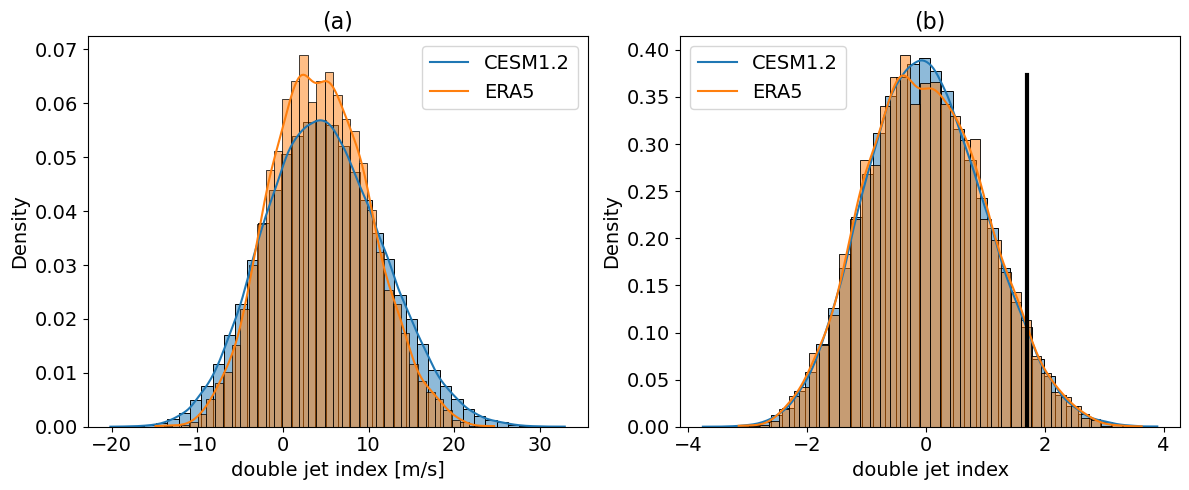}
        \caption{(a) Histograms and kernel density estimation of the daily double jet index for CESM1.2 (blue) and ERA5 (orange). (b) The deseasonalised and standardized histograms. The vertical lines indicate the thresholds corresponding to the 5\% most extreme events chosen to define a double jet in \cref{fig:vertical_zonalwind_cesm,fig:vertical_zonalwind_era}.}
        \label{fig:dji-hist-ERA-CESM-new}
    \end{figure}

    In this section, we use the double jet index defined in \cref{subsec:dji_def} to identify double jet states in the CESM1.2 control run and in ERA5. \Cref{fig:dji-hist-ERA-CESM-new}a shows the probability density function of the daily double jet index over the June-July-August period (JJA), for CESM1.2 (blue) and ERA5 (orange). We see that in both data sets, the mean value is positive, around 5 m s$^{-1}$. However the variance is different and the two distributions differ in the tails. \Cref{fig:dji-hist-ERA-CESM-new}b shows the same histograms, after removing the climatological mean and dividing by the standard deviation. We see that the tail of the distribution is similar between CESM1.2 and ERA5 after the rescaling. This suggests that defining double jet states in the two datasets based on a percentile threshold of the index would identify comparable events.

\begin{figure}[t]

        \includegraphics[width=1\textwidth]{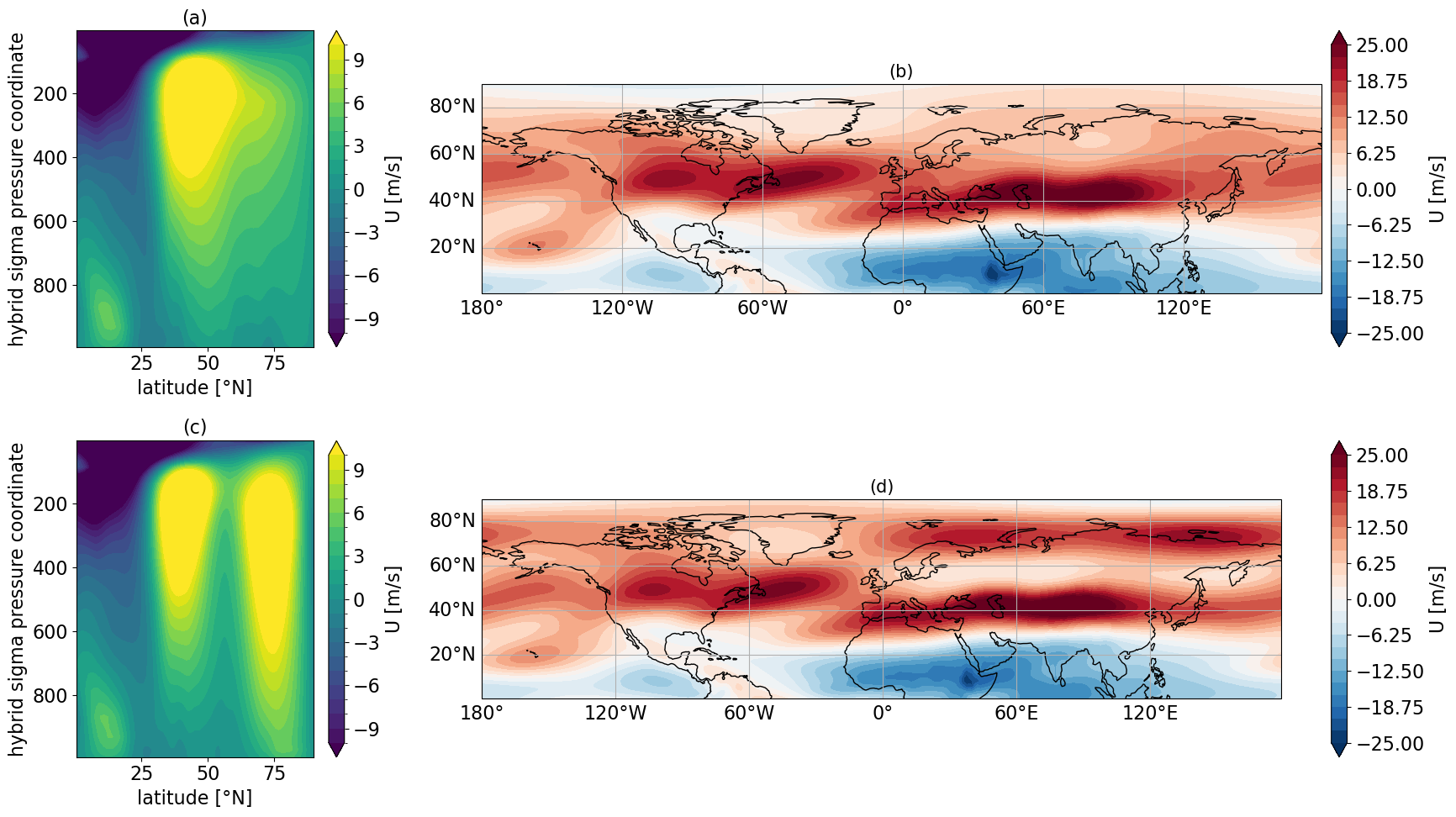}
        \caption[Vertical profile of the daily summer zonal wind and zonal wind map for CESM1.2.]{(a) Vertical profile of the daily summer zonal wind $U$ for the climate model CESM1.2 (1000 years), longitudinally averaged over the Eurasian sector (10$^\circ$W to 180$^\circ$E). We considered the months of June, July and August  - (b) Zonal wind $U$ climatology for the CESM1.2 climate model (1000 years) for the months of June, July and August, (c) - (d) same variable as for (a) - (b) but when selecting days corresponding to the 5\% most extreme values of the double jet index. There is a clear sign of a second separate jet appearing at high latitudes.}
        \label{fig:vertical_zonalwind_cesm}
    \end{figure}

\begin{figure}[t]

         \includegraphics[width=1\textwidth]{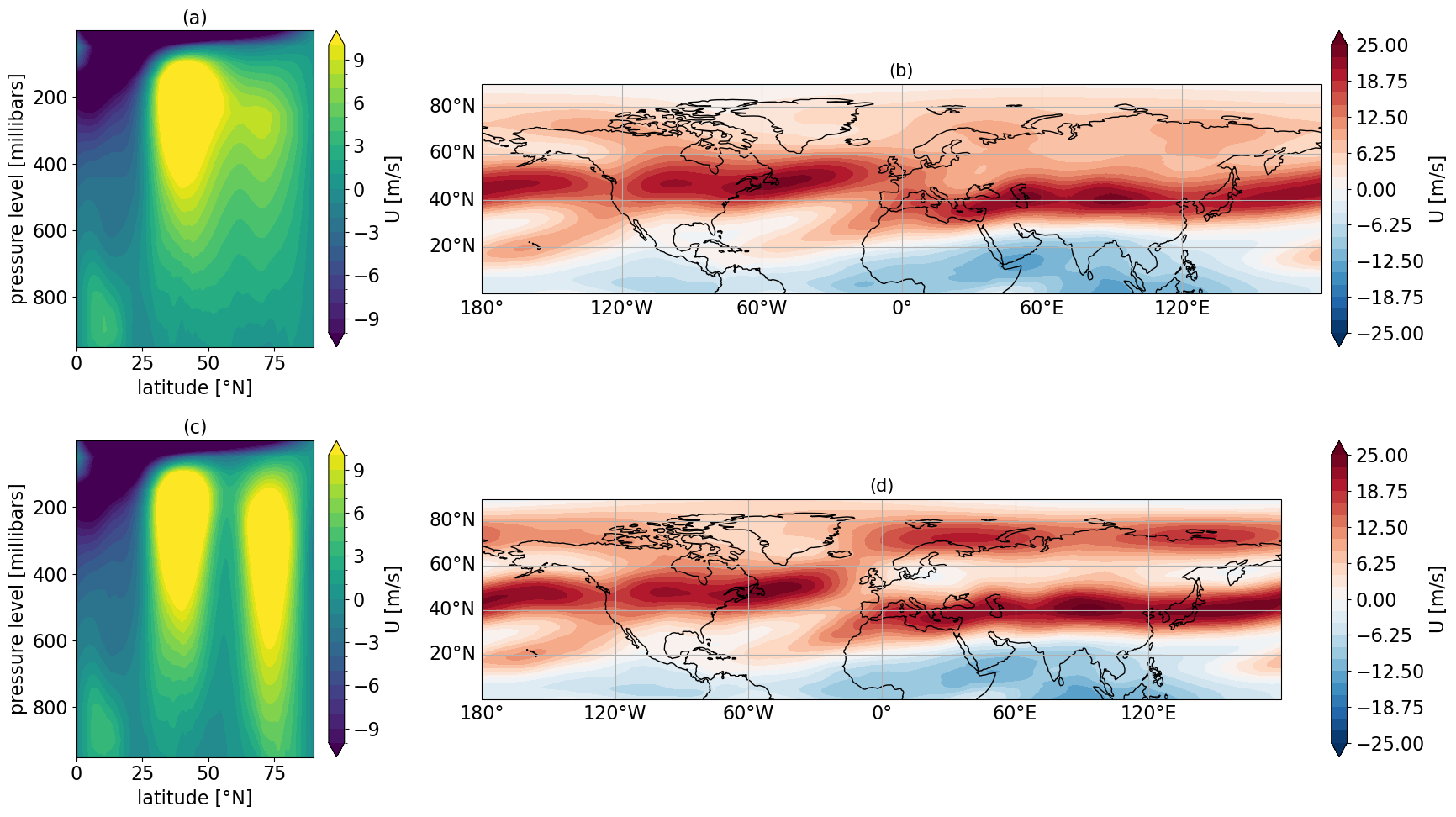}       
         \caption[Vertical profile of the daily summer zonal wind and zonal wind map for ERA5.]{(a) Vertical profile of the daily summer zonal wind $U$ for the reanalysis dataset ERA5, longitudinally averaged over the Eurasian sector (10$^\circ$W to 180$^\circ$E). We considered the months of June, July and August  - (b) Zonal wind $U$ climatology for the reanalysis dataset ERA5 for the months of June, July and August, (c) - (d) same variable as for (a) - (b) but when selecting days corresponding to the 5\% most extreme values of the daily double jet index. There is a clear sign of a second separate jet appearing at high latitudes.}
         \label{fig:vertical_zonalwind_era}
    \end{figure}

    We thus define a double jet state as one for which the amplitude $D$ exceeds the 95th percentile (marked with vertical lines in \cref{fig:dji-hist-ERA-CESM-new}b). This corresponds to a double jet index value of 16.3m s$^{-1}$ and 14 m s$^{-1}$ for CESM1.2 and ERA5 respectively. 
    \Cref{fig:vertical_zonalwind_cesm} (a) and (b) show respectively the CESM1.2 JJA climatology of the latitude - height structure of the zonal mean $U$  longitudinally averaged over the Eurasian sector (10$^\circ$W to 180$^\circ$E) and the zonal wind vertically averaged over 192 and 313 hybrid sigma pressure coordinate (the same levels as for the index $D$). We see a main single jet in mid latitudes. The presence of weak second peak in the climatological average is consistent with the fact that the climatological mean of the double jet index is positive~(\Cref{fig:dji-hist-ERA-CESM-new}a). 
    \Cref{fig:vertical_zonalwind_cesm} (c) and (d) show the same but restricting to days identified as double jet days. We can clearly see the appearance of a double jet structure in the cross section, corresponding to the emergence of a band of strong westerlies north of about 65°N extending over the entire Eurasian sector.  
    \Cref{fig:vertical_zonalwind_era} shows the same results for ERA5. The similarity between the CESM1.2 and ERA5 results suggests that 1) our double jet index identifies events that yield consistent double jet composites across reanalysis and model, and 2) CESM1.2 captures a summer Eurasian double jet state comparable to that in ERA5. Indeed, the temperature composite maps are consistent with those obtained for ERA5 in \cite{rousi_accelerated_2022}, where double jet events were identified using a machine-learning approach.

\begin{figure}[t]
        \includegraphics[width=1\textwidth]{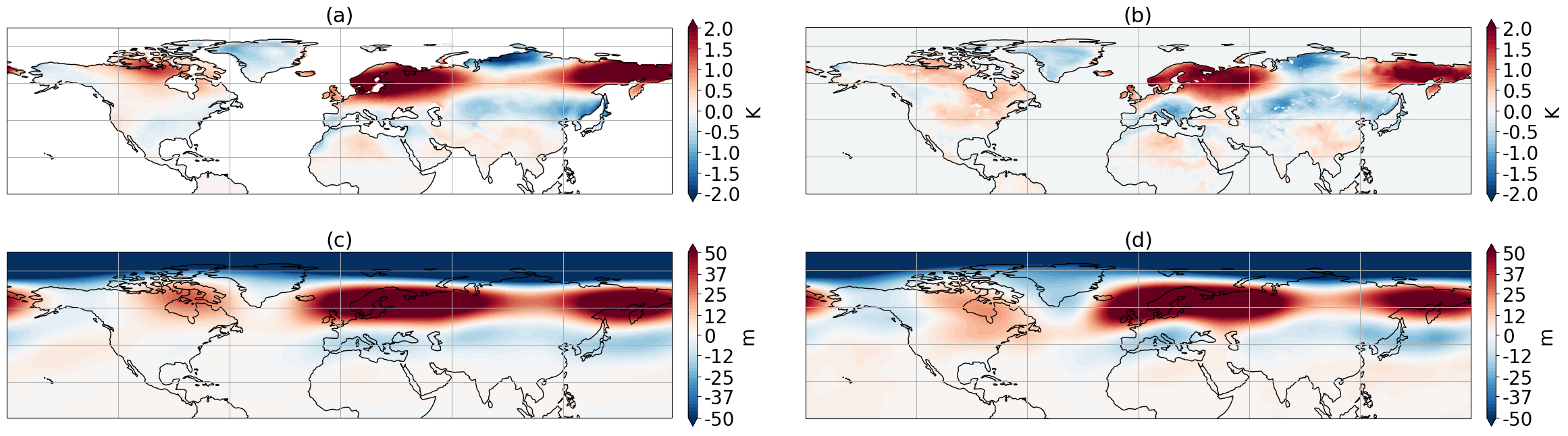}
        \caption{Composite maps for (a) $T_{2m}$ and (c) $Z_{500}$ anomalies for 5\% most extreme double jet days for CESM1.2 control run, (b) - (d) same but for ERA5.}
        \label{fig:composite_maps_cesm-era}
    \end{figure}

    To examine the relation of the double jet to the large scale circulation and heatwaves, we next composite the $T_{2m}$ and $Z_{500}$ during extreme double jet days, in CESM1.2 and ERA5 (\cref{fig:composite_maps_cesm-era}, the respective standard deviation is in 
    \cref{fig:std_maps_cesm-era}).
    We can see that these states are characterized by positive temperature anomalies and associated anomalous large-scale highs over three regions: North Canada, Scandinavia and East Russia, alongside a negative geopotential height anomaly over the Arctic. 
    This $Z_{500}$ anomaly pattern, characterized by a pronounced negative center over the central Arctic and predominantly positive anomalies over the mid-latitudes, resembles the summertime NAM pattern shown in \cite{ogi_summertime_2004}. In particular, their Fig. 2(e), which shows 500-hPa geopotential height regressed on the summer (June–July) SV NAM index, features a strong cyclonic circulation anomaly over the Arctic and positive anomalies across the mid-latitudes, which closely resembles our composite map (our \cref{fig:composite_maps_cesm-era}). The connection between the Northern Annular Mode and the double jet is discussed further in the next sections. In addition, the midlatitude high-pressure anomalies are distributed according to a wavenumber 3 pattern similar to the one identified in \citep{ragone_rare_2021} to be associated with extreme heatwaves over France and Scandinavia.

   \subsection{Persistent double jet states and heatwaves} 
    \label{sec:dji_hw}
    
    The composite maps in figure \cref{fig:composite_maps_cesm-era} do not take into account the persistence of the double jet structures. In order to test whether persistent double jet states are related to more severe heatwaves, we computed the accumulated double jet index as described in \cref{subsec:dji_def,subsec:hw_def} for durations ranging from 1 to 90 days (by taking the temporal average of daily data over a fixed duration). Then we computed an accumulated heatwave index following the same procedure, over the three areas identified in \cref{fig:composite_maps_cesm-era}, respectively North Canada (50°N-80°N,130°W-60°W), Scandinavia (55°N-72°N,3°E-50°E) and East Russia (50°N-75°N,120°E-180°E), shown in 
    \cref{fig:hw_regions}
    . For each duration we then defined double jet and heatwave days (separately for the three regions) based on the $95^{th}$ percentile of the corresponding index. 

\begin{figure}

        \centering
        \includegraphics[width=0.5\textwidth]{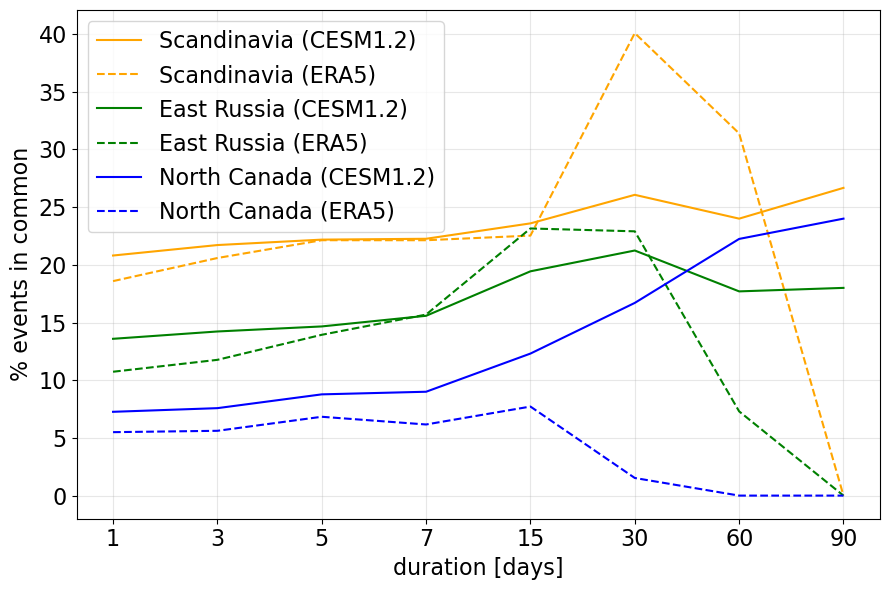}
        \caption[Percentage of days in common between double jet and heatwave events ]{Percentage of days in common between double jet (defined in \cref{subsec:dji_def}) and heatwave events (defined in \cref{subsec:hw_def}) as a function of the duration in days, for different regions: North Canada (blue), Scandinavia (orange) and East Russia (green), for CESM1.2 (solid line) and ERA5 (dashed line). The spatial areas are shown in 
        \cref{fig:hw_regions}.
        Both double jet and heatwaves are defined as events based on the $95^{th}$ percentile of the distribution of both indexes, i.e. we consider as double jet or heatwaves the 5\% most extreme events in our datasets.}
        \label{fig:hw_dji_cesm_era}
        
    \end{figure}

    \Cref{fig:hw_dji_cesm_era} shows for all regions the percentage of events that are both double jet and heatwave days as a function of the accumulation period in the definition of the indexes, for both CESM1.2 and ERA5. We observe that in CESM1.2 for all three regions there is a clear increase with the accumulation period of the percentage of days in common between double jet states and heatwaves. This means that long lasting double jet events are more strongly correlated to long lasting heatwave events than their short-lived counterparts. In ERA5 the relation is less clear. For Canada there is no apparent increase in the percentage of days in common, while for Scandinavia and East Russia there is an increase up to an accumulation period of about 30 days, after which there is a drop. However, we need to stress that, given the short length of the ERA5 dataset, the uncertainties are very large, in particular for long lasting events (note that, since we consider only the JJA period, when taking an accumulation period of 90 days we remain with just one value per year). 

    \subsection{Persistent double jet associated to strong polar cyclonic anomalies}
    \label{sec:dj_am}
\begin{figure}

        \includegraphics[width=1\textwidth]{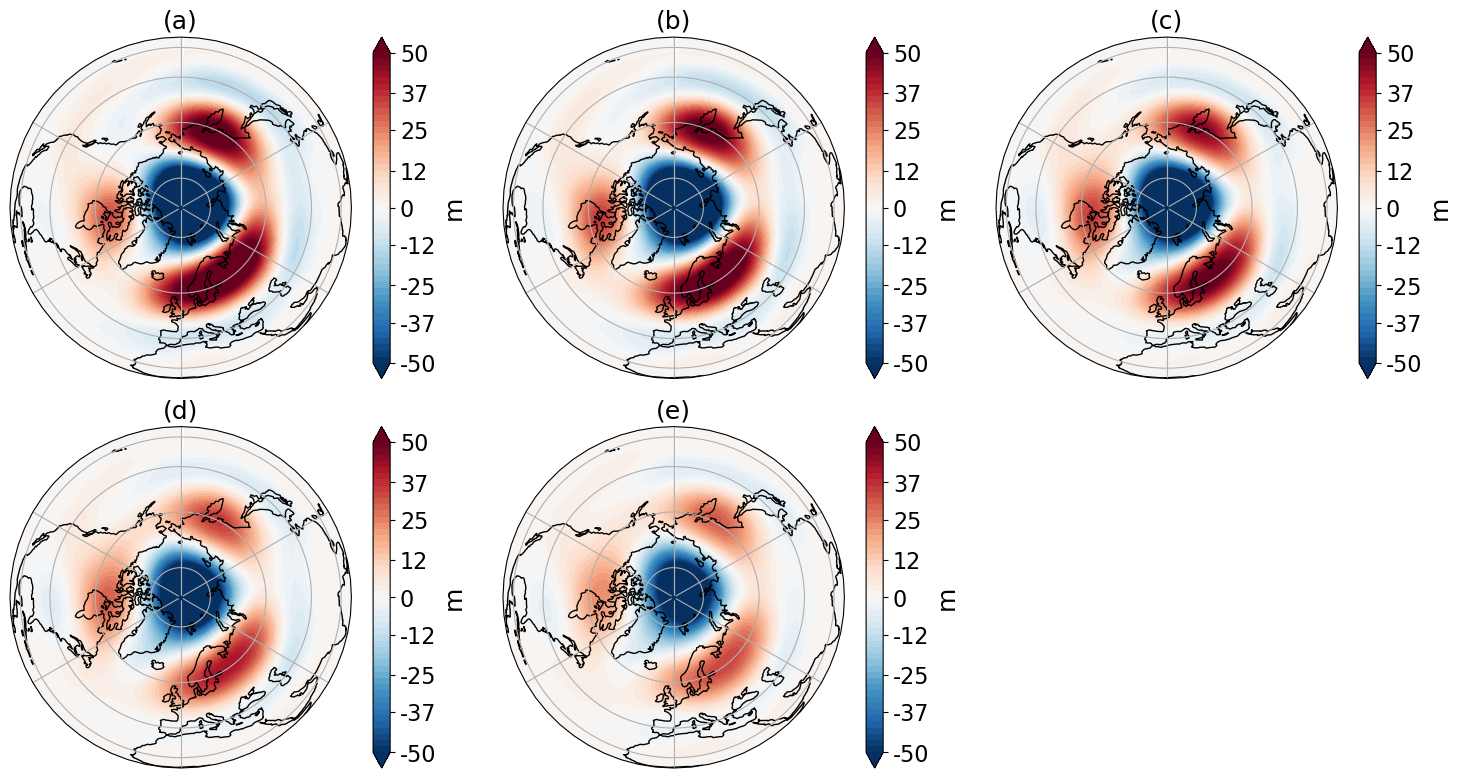}
        \caption{Composite maps for CESM1.2 of $Z_{500}$ anomalies for 5\% most extreme double jets, for different durations of the double jet events: (a) 1-day, (b) 7-day, (c) 14-day, (d) 30-day, (e) 90-day. No matter the duration of the double jet events, their are associated with a strong low pressure system over the polar region and strong anticyclonic anomalies over the midlatitudes.}
    \label{fig:zg500-ortho}
    \end{figure}

    In the previous section we investigated the link between persistent double jets and heatwaves. In this section, we focus on the link between double jets and persistent \SI{500}{\hecto \pascal} geopotential height anomaly ($Z_{500}$). Similarly to what has been done in the previous section, we computed the accumulated double jet index as described in \cref{subsec:dji_def} for durations ranging from 1 to 90 days (by taking the temporal average of daily data over a fixed duration), we define a double jet state as one for which the amplitude $D$ exceeds the 95th percentile and we then composite the $Z_{500}$. Results for 1-day events were shown already in \cref{subsec:dji_cesm_era}. Here we extend the analysis up to persistent seasonal events. The composites are shown in \cref{fig:zg500-ortho} for CESM1.2 and in 
    \cref{fig:zg500-ortho-era} 
    for ERA5. From these maps, we see that no matter the duration of the double jet events, they are associated with a wave 3 pattern in the $Z_{500}$ at the midlatitudes and a strong low pressure wave zero on the polar cap. As we increase the duration of the double jet events, we observe a reduced amplitude in the $Z_{500}$ pattern. 

\begin{figure}

        \includegraphics[width=1\textwidth]{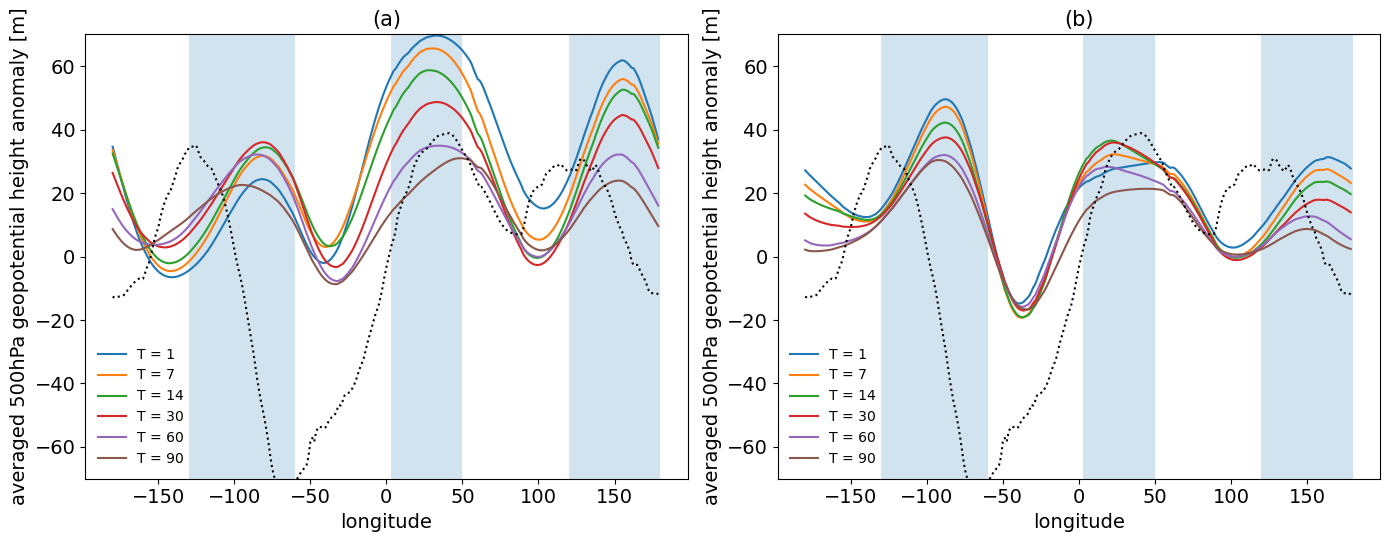}
        \caption{$Z_{500}$ anomalies (for CESM1.2) averaged over the latitudinal band 55°N-70°N, conditioned on the 5\% most extreme a) double jet events and b) positive Annular Mode events, with the events defined based on different time average periods ($T$). Also shown for reference is the corresponding climatological stationary wave (dotted line). Blue-shaded regions highlight the longitudinal extensions of the three regions used to define a heatwave, starting from the left: North Canada, Scandinavia and East Russia.}
        \label{fig:averaged_zg_T}
    \end{figure}

    However, despite the reduced amplitude of the wave, it becomes more global as the double jet events span over an entire season. To check that, we take a latitudinal average over the sector 55$^\circ$N-70$^\circ$N of the composite maps in \cref{fig:zg500-ortho}. The result is shown in \cref{fig:averaged_zg_T}a, alongside the \SI{500}{\hecto \pascal} climatological stationary wave for reference (dotted line). We see that the Scandinavian peak of double jet related anomaly pattern interferes positively with the climatological stationary waves, while the Siberian peak is more of an eastward shift, and the Canadian peak mostly a negative interference of the climatological positive stationary wave peaks. The phasing relative to climatology being strongest for the Scandinavian peak and weakest for the Canadian one might explain there being the largest percentage co-occurrence between the double jet and heatwaves for the former and the smallest co-occurrence for the latter. 

    As we already mentioned in \cref{sec:results_dji} for 1-day events, the $Z_{500}$ composite map strongly resembles the summer NAM investigated by \citep{ogi_summertime_2004} (see their figure 2e). In particular, \citep{ogi_summertime_2004} found that double jets appears in presence of a large positive phase of the summer NAM. In order to test this, we compute the same index as in their work. The only difference is that we computed it using the \SI{500}{\hecto \pascal} geopotential height anomaly, instead of averaging between \SI{1000}{\hecto \pascal} and \SI{200}{\hecto \pascal}, because of data availability. However, as mentioned by the authors in that work, results are not sensitive to this choice. 
    \Cref{fig:averaged_zg_T}b shows the composites of geopotential height when this index averaged over different time periods is above the 95th percentile. Comparing to the corresponding double jet index pattern (\cref{fig:averaged_zg_T}a), we see that both anomalies project onto the same longitudinal centers, suggesting a strong connection between them. Specifically, the two patterns are most similar for the 90-day averages.     
    
\begin{figure}
        \centering
        \includegraphics[width=0.7\linewidth]{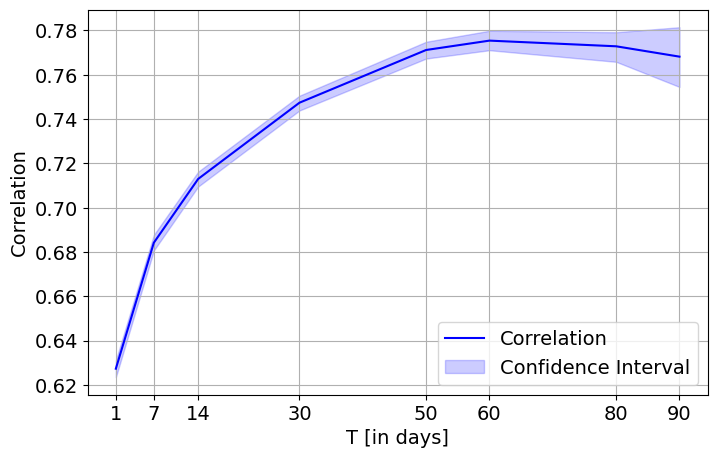}
        \caption{Correlation between the double jet index, defined in \cref{subsec:dji_def} and the Annular Mode index (see \cref{sec:dj_am}) for CESM1.2, which follows closely the one in \cite{ogi_summertime_2004}. Shades indicate confidence interval obtained with a bootstrap test with a 95\% confidence.}
        \label{fig:corr_dji_am}
    \end{figure}

    \Cref{fig:corr_dji_am} shows the Pearson correlation coefficient between the double jet and the Annular Mode indexes. We see that high values of the double jet index and high values of the Annular Mode are moderate to strongly correlated. This shows that states with clear double jet structures are associated with strong low pressure systems over the polar cap. Moreover, as the correlation gets stronger the longer events are considered, this plot shows that persistent double jets are associated with persistent Annular Mode and wave 3 pattern. This further motivates the study of persistent double jet summers with the aid of a rare events algorithm as a technique to mitigate the high computational cost associated with climate simulations. 

   \subsection{Importance sampling of extreme double jet summers}
    \label{sec:rea_results}
    
\begin{figure}
        
        \includegraphics[width=1\textwidth]{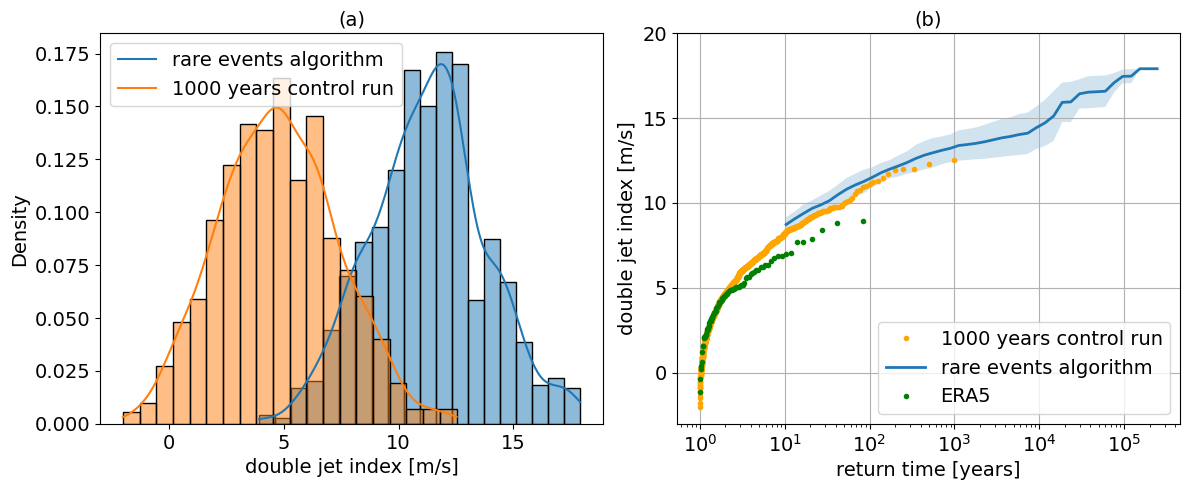}
        
        \caption[(a) Histograms and kernel density estimation of the seasonal JJA double jet index in the 1000 years control run of CESM1.2 (orange) and with the rare events algorithm (blue), (b) Return time curves for the control run of CESM1.2 (1000 years) in orange and the one obtained with 10 runs of the rare event algorithm in blue, and ERA5 in green.]{(a) Histograms and kernel density estimation of the seasonal JJA double jet index in the 1000 years control run of CESM1.2 (orange) and with the rare events algorithm (blue) with $k = 0.01 (m s^{-1})^{-1}(day)^{-1}$. (b) Return time curves for the control run of CESM1.2 (1000 years) in orange and the one obtained with 10 runs of the rare event algorithm in blue, and ERA5 in green. The dark blue line represents the ensemble mean, while the shadow blue region corresponds to one standard deviation. 
        \label{fig:rt-gktl-k001}
    }
    \end{figure}

    The persistent double jet structures analyzed in the previous sections are defined as events in the upper 5\% of the double jet index distribution and further constrained by long persistence, which makes them extremely rare. Even in the CESM1.2 data, statistical uncertainties remain large for the most extreme and long-lasting events.
    Here we apply a rare events algorithm to new simulations with CESM1.2 as described in \cref{subsec:rea}, in order to oversample model trajectories characterized by persistent double jet states. \Cref{fig:rt-gktl-k001}a shows the histograms and kernel density estimations of the double jet index, averaged over 90 days, for the control run (orange) and simulations with the rare events algorithm (blue). We can see that the algorithm successfully performs an importance sampling of the distribution, by sampling more efficiently the upper tail and even events which were not present in the control distribution. The typical value for the double jet index in the resampled distribution is 11.3 m s$^{-1}$, corresponding to the 100-year return level, which is very much rare in the control run. 
    
    In \cref{fig:rt-gktl-k001}b, we plot the return time of the 90-day double jet index for both the control and the rare events algorithm simulations, and ERA5. A description of how these curves are obtained can be found in 
    \cref{supmat:rt_curves}.
   The black curve is evaluated using the control run, while the green one using ERA5 reanalysis data. The blue curve is obtained by computing return time curves for each of the 10 ensemble experiments with the rare events algorithm, and then averaging over the 10 results. The shaded area indicates the uncertainty computed as one standard deviation of the sample average (see 
    \cref{supmat:rt_curves} 
    and \citep{ragone_computation_2018} and \citep{ragone_rare_2021} for a more detailed description of the procedure). We can see that thanks to the rare events algorithm we are able to sample events with return times up to $10^5$ years, with a computational cost of $10^3$ years of simulations.

\begin{figure}[t]
            \includegraphics[width=1\textwidth]{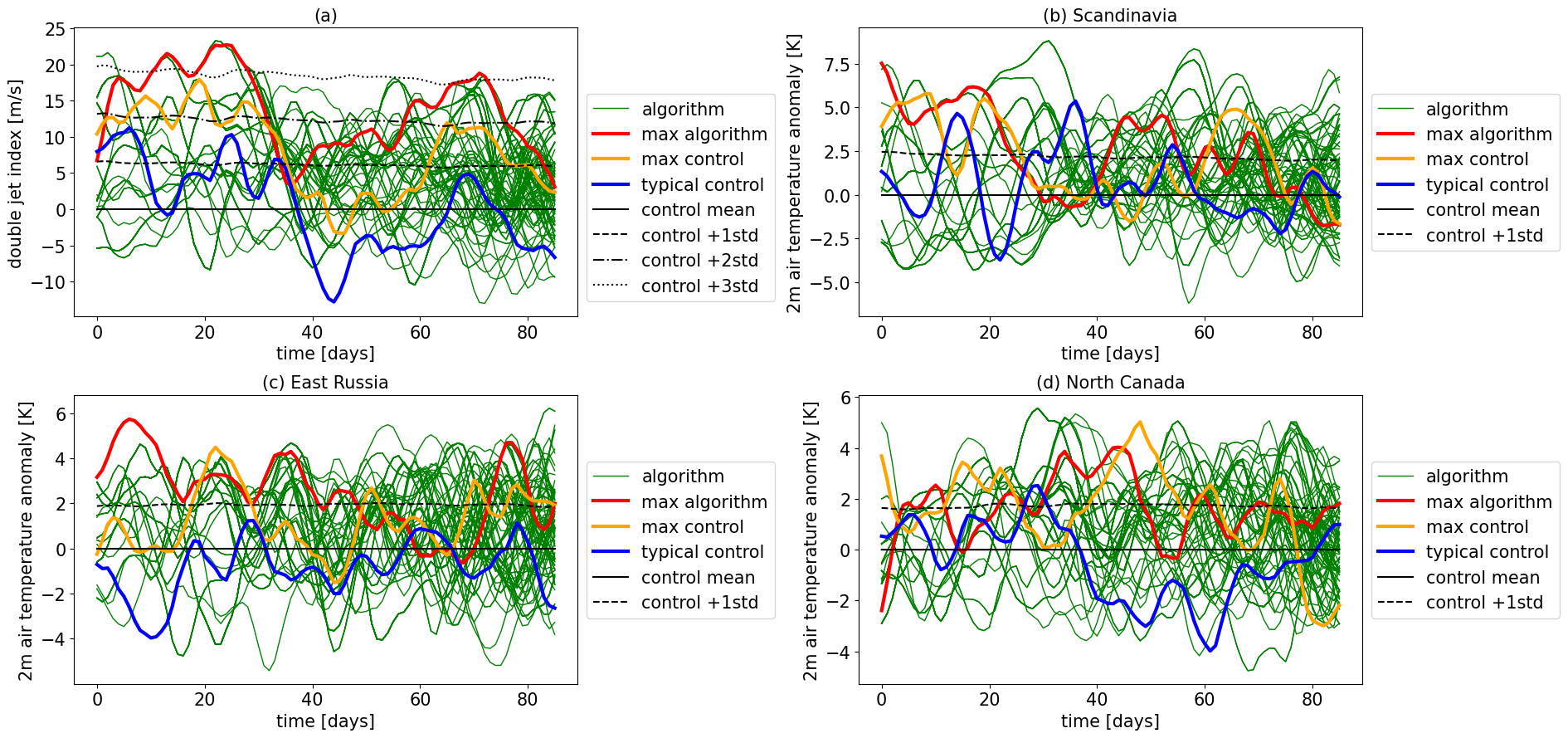}
        \caption{ (a) Dynamics of the double jet index for one experiment of the GKLT algorithm. The green are all the 100 trajectories in that experiment and the red one is the one with the highest seasonal double jet index in that experiment. The yellow one is the trajectory with the highest seasonal double jet index in the control, while the blue is a trajectory with a seasonal double jet index around the mean. Note that we removed the mean to have a fair comparison with the other panels in the figure. The time on x-axis starts on 1st of June until the 29th of August. (b)-(d) Dynamics of the 2 meter air temperature anomaly for the same experiment as in panel (a), for the three different heatwave regions. The red and yellow trajectories are selected according the highest seasonal double jet in the rare events algorithm experiment, while the blue one a trajectory with a mean double jet index value in control run. The time on x-axis starts on 1st of June until the 29th of August.}
        \label{fig:dji_t_trajs}
    \end{figure}

    \Cref{fig:dji_t_trajs}a shows the temporal evolution of the double jet index for trajectories of one experiment with the rare events algorithm. The red trajectory corresponds to the summer with the highest seasonal double jet index reached in that experiment of the rare events algorithm, while in green we show all the trajectories for that experiment. The yellow one is the strongest summer with the highest seasonal double jet index for the control run. Finally, the blue one, named as the typical, has a seasonal double jet value which fluctuates around its mean value. We can see that for typical trajectories the index fluctuates around the climatology on a scale of 5-10 days, corresponding to the time scales of the synoptic fluctuations and consistently with its autocorrelation function. Large values of the 90-day double jet index are obtained with large but more importantly persistent excursions of the index, that features an average value systematically above 2 standard deviation the climatology, with 5-10 days fluctuation overimposed on the mean state. This indicates that these events are truly persistent throughout the entire summer season and correspond to the activation of a slow mode of variability of the atmosphere, with the regular synoptic variability embedded within it. In the same figure, \cref{fig:dji_t_trajs}b-d show the temporal evolution of the \SI{2}{\meter} air temperature for the summers with the highest seasonal double jet index both in the rare events algorithm experiment and the control run, namely the red and orange trajectories, for three regions ((b) Scandinavia (c) East Russia (d) North Canada). Moreover, similarly to \cref{fig:dji_t_trajs}a, we also show a trajectory selected on seasonal double jet that fluctuates around its mean and all the trajectories of that experiment in green. In all zones, large values of the 90-day double jet index are associated with great excursions of the \SI{2}{\meter} air temperature above 1 standard deviation of the climatology, which persist for large periods, up to a season for the East Russia.

\subsection{Teleconnection patterns for double jet summers}
\label{sec:rea_teleconection}
\begin{figure}
    \includegraphics[width=1\textwidth]{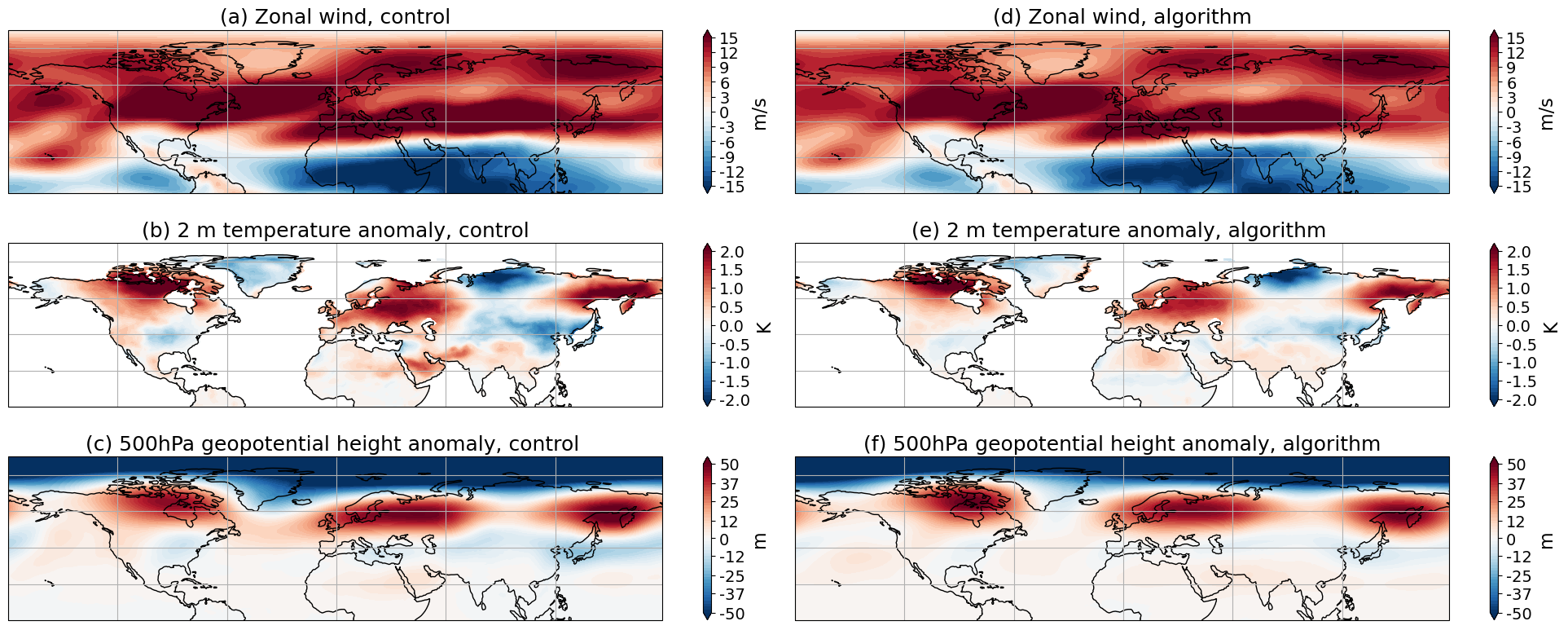}
    \caption[Composite maps for $U$, $T_{2m}$ and $Z_{500}$ for 100 years return time.]{Composite maps for (a) $U$, (b) $T_{2m}$ and (c) $Z_{500}$ for 100 years return time of double jet index averaged over the summer months of June, July and August for CESM1.2 control run (average over 10 maps); (d) -(e) same but for the rare events algorithm.}
    \label{fig:composite_maps_ctrl_gktl}
\end{figure}

We analyse the dynamics during summers characterized by values of the 90-day double jet index with return times larger than 100 years (denoted as 100-year double jet summer or events from now on). In \cref{fig:composite_maps_ctrl_gktl} we show the composite maps for the zonal wind $U$ (vertically averaged between 192 and 313 hybrid sigma pressure coordinate), the \SI{2}{\meter} temperature anomaly ($T_\text{2m}$) and \SI{500}{\hecto\pascal} geopotential height anomaly ($Z_\text{500}$) for values of the 90-day double jet index larger than its 100-year return level (11.3 m s$^{-1}$), for the control run (left column) and the rare events algorithm (right column). Starting from the top row, we see the appearance of a poleward second branch over Eurasia. Associated with this configuration, we observe the emergence of three zones with positives $T_\text{2m}$ and a wave number 3 in $Z_\text{500}$, consistently with the previous analysis. 

In order to test the statistical significance of these patters, we show in 
\cref{fig:t-values-ctrl-gktl} 
the t-values obtained from a Student t-test for the different fields. Details of this statistical test are given in 
\cref{supmat:ttest}.
For the control run, the zonal wind $U$ has the largest amount of significant zones, which is not surprising given that is it strictly connected to the double jet definition. For the other two fields, we observe some significant zones in North Canada, East and North Europe and over East Russia, and over the Polar Circle for $Z_\text{500}$. The right columns shows the t-values for the rare event simulations. Given the larger amount of sampled events in the rare event algorithm simulations, the significant zones are wider. The algorithm results are globally significant, except in some areas, such as in the United States of America. The results of the algorithm enables to prove the existence of significant teleconnections associated with a double jet summer, particularly in North Canada, Europe and East Russia. While this pattern is present for $T_\text{2m}$ in the control simulation, although with limited significant zones, for the $Z_\text{500}$ the wave number is not clear in the control simulation, especially over the Atlantic Ocean. 

The teleconnection pattern is similar to the one shown in \citep{ragone_computation_2020}, where the authors used the same climate model and rare event algorithm presented here  for the sampling of warm summers in both France and Scandinavia. The patterns are significantly similar for Scandinavian heatwaves and a careful comparison with the processes discussed in \citep{ragone_computation_2020} might be an interesting future pathway. We also mention that another study from the same authors \citep{ragone_computation_2018} which used the same rare events algorithm but a different climate model and area, produced similar results in terms of teleconnection patterns. These results support the idea that teleconnection at subseasonal time scales and the associated large scale dynamics corresponding to a wave number 3 are robust features. 

\begin{figure}
    \centering

    
    
    
    

    \includegraphics[width=0.75\textwidth]{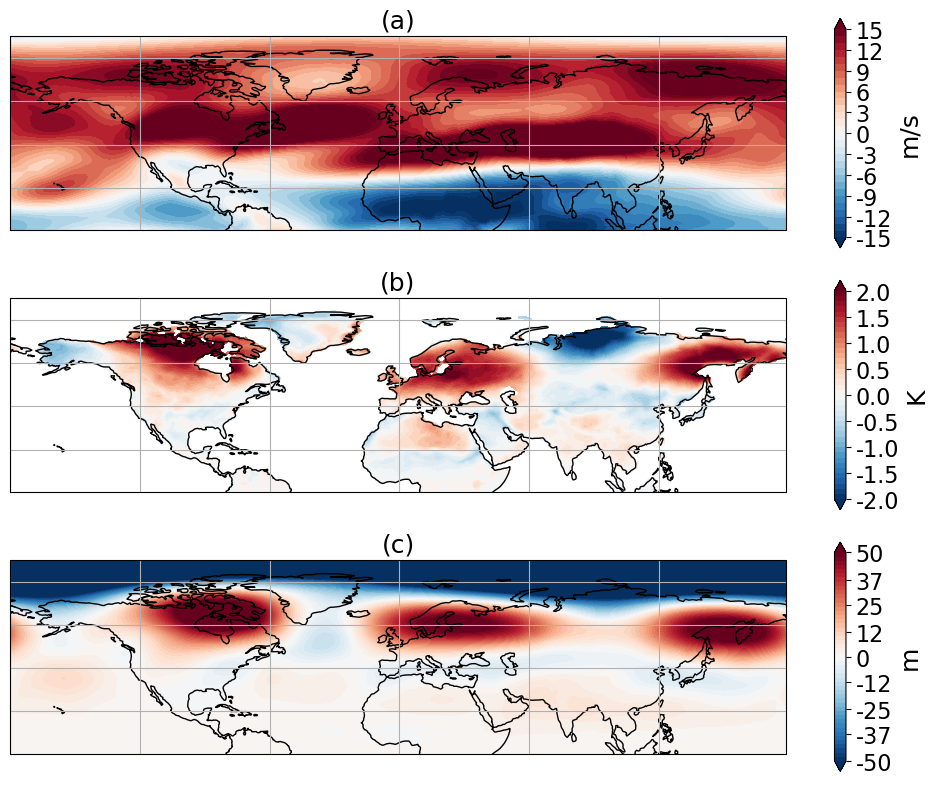}
    \caption[Composite maps for $U$, $T_{2m}$ and $Z_{500}$ for 1000 years return time.] {Composite maps for (a) $U$, (b) $T_{2m}$ and (c) $Z_{500}$ for 1000 years return time of double jet index averaged over the summer months of June, July and August for the rare events algorithm.}
    \label{fig:composite_maps_gktl1000}
\end{figure}

The rare events algorithm allows to sample events which are unseen in the control run. In figure \cref{fig:composite_maps_gktl1000} we plot the dynamical fields of double jet summers with a return time larger than 1000 years (denotes as 1000-year double jet summer or events from now on). It is interesting to notice that the patterns of the composite maps of 1000-year double jet summer are similar to the ones of the 100-year double jet summer, for all the fields, with an obvious increasing in the amplitude. This is consistent with previous results of the application of this class of algorithms to other climate phenomena \citep{ragone_computation_2018, ragone_rare_2021, cozian_computing_2023, wouters_rare_2023, cini_simulating_2024, sauer_extremes_2024, lestang_numerical_2020,le_priol_using_2024}. 

\section{Conclusions}
\label{sec:conclusions_dji}

In this paper we have investigated the relation between double jet structures in the Northern hemisphere atmosphere and the occurrence of heatwaves over Scandinavia, East Russia and Northern Canada. We introduced an index based on the upper troposphere zonal wind that successfully identifies double jet states on different time scales. Based on ERA5 data and a stationary run with CESM1.2 we showed that there is a significant percentage of overlapping days between double jet and heatwaves, and that this percentage increases for more persistent events. Then we applied a rare event algorithm to ensemble simulations with CESM1.2, improving the statistics of very rare persistent double jet states. For all datasets, we find that extreme double jet states, for all averaging periods between a day to the whole summer, are characterised by three centers of extreme high surface temperature and 500hPa geopotential height anomalies,
alongside a strong low pressure over the Arctic. The geopotential height anomaly
pattern is consistent with both a positive Northern Annular Mode~\citep[consistent with][]{ogi_summer_2005} and the quasi-wave-3 pattern found by~\citep{ragone_rare_2021} to be connected to extreme heatwaves.

We further note that differently from previous applications of the rare event algorithm, in this study we sample directly persistent atmospheric states which are prone to support extreme events. We stress that this was not the case in previous works, where for example the same rare events algorithm was applied for sampling of extreme warm summers and the atmospheric states were retrieved afterwards \citep{ragone_computation_2018,ragone_rare_2021}. 

An analysis of composite maps of \SI{2}{\meter} temperature anomaly field and of the \SI{500}{\hecto\pascal} geopotential height anomaly, conditioned over the occurrence of a 1-in-100 years double jet summer, reveals a significant wave number 3 pattern, with positive temperature and anticyclonic anomalies over North Canada, North Europe/Western Russia and Eastern Russia. Similar teleconnection patterns and with the same wave number were found by \citep{ragone_rare_2021} using the same climate model and rare events algorithm as used in this study for simulating warm summer over the Scandinavia peninsula. An interesting future perspective is to compare both studies to assess more quantitatively the role of double jet in Northern European heatwaves. In the literature, double jet structures have been analysed in correspondence with waves with higher wave numbers (typically between 5 and 7) \citep{kornhuber_extreme_2019,coumou_quasi-resonant_2014,petoukhov_role_2016}. A natural follow-up would be to investigate the appearance of these wave numbers in the context of double jet.

The composite patterns obtained with the algorithm for very large return times resemble in geographical structure the ones with return times higher than 100 years. This indicates that rare, seasonally persistent double jet states are realised with a consistent dynamics, associated with a phased locked wavenumber 3 hemispheric pattern. This result is consistent with our previous finding \citep{ragone_computation_2018,ragone_rare_2021} and with the concept of dynamical typicality explored in \citep{lucarini_typicality_2023,galfi_fingerprinting_2021,noyelle_investigating_2024}. according to which extreme fluctuations of a given large-scale observable are realized through dynamically similar trajectories that mainly differ in amplitude rather than in spatial structure. A complementary statistical interpretation has also been recently proposed to motivate this behavior. In \citep{mascolo_gaussian_2025}, the authors devised a framework which correctly captures the scaling of the composite maps with the threshold level used to define the extreme events. With simple but meaningful assumptions between the weather fields which characterize the event and the metrics used to define the event, the authors found an analytical expression for this scaling. In that work, the framework was applied to the analysis and forecasting of heatwaves over France. An interesting future perspective could be to apply the same methodology to the double jet index presented here.

In this study we have used a prescribed SST and sea ice setup for the model. This setup excludes two-way ocean–atmosphere coupling and limits for example the role of tropical SST-driven teleconnections like ENSO. However, the very fact that we observe this dynamical states is an indication that they arise primarily from internal atmospheric variability and its interactions with land/sea-ice boundary conditions, rather than from coupled ocean–atmosphere feedbacks. In a fully coupled framework, these feedbacks could alter the intensity, duration, and frequency of double jet states and their association with heatwaves, potentially more so under climate warming, as it was suggested in \cite{rousi_accelerated_2022}.
\clearpage
\acknowledgments
This project was provided with computing and storage resources by GENCI at TGCC on the partition ROME of supercomputer Joliot Curie thanks to the grant AD010110575R2. V. Mascolo has received funding from the European Union’s Horizon 2020 research and innovation program under the Marie Skłodowska-Curie grant agreement 956396 (EDIPI). N. Harnik acknowledges the Israeli Science Foundation grant number 2466/23. The authors thank the computer resources provided by the Centre Blaise Pascal at ENS de Lyon. We are grateful to Emmanuel Quemener for his help with the platform. V. Mascolo thanks B. Cozian for his help with running CESM1.2.

%
%
\datastatement
Data from the reanalysis dataset ERA5 \citep{hersbach_era5_2020}, publicly available at \url{https://www.ecmwf.int/en/forecasts/dataset/ecmwf-reanalysis-v5} are used in this study. We also use data from a long simulation (1000 years) of the CESM1.2 climate model available at \citep{ragone_rare_2021}. Data from the rare events algorithm simulations are publicly available at \url{https://doi.org/10.5281/zenodo.15593945}.
%

\appendix




\appendixtitle{Supplementary information}

%



\subsection{Detrending of ERA5 dataset}
    \label{supmat:era}
\begin{figure}[h]
        \centering
        \includegraphics[width=0.6\textwidth]{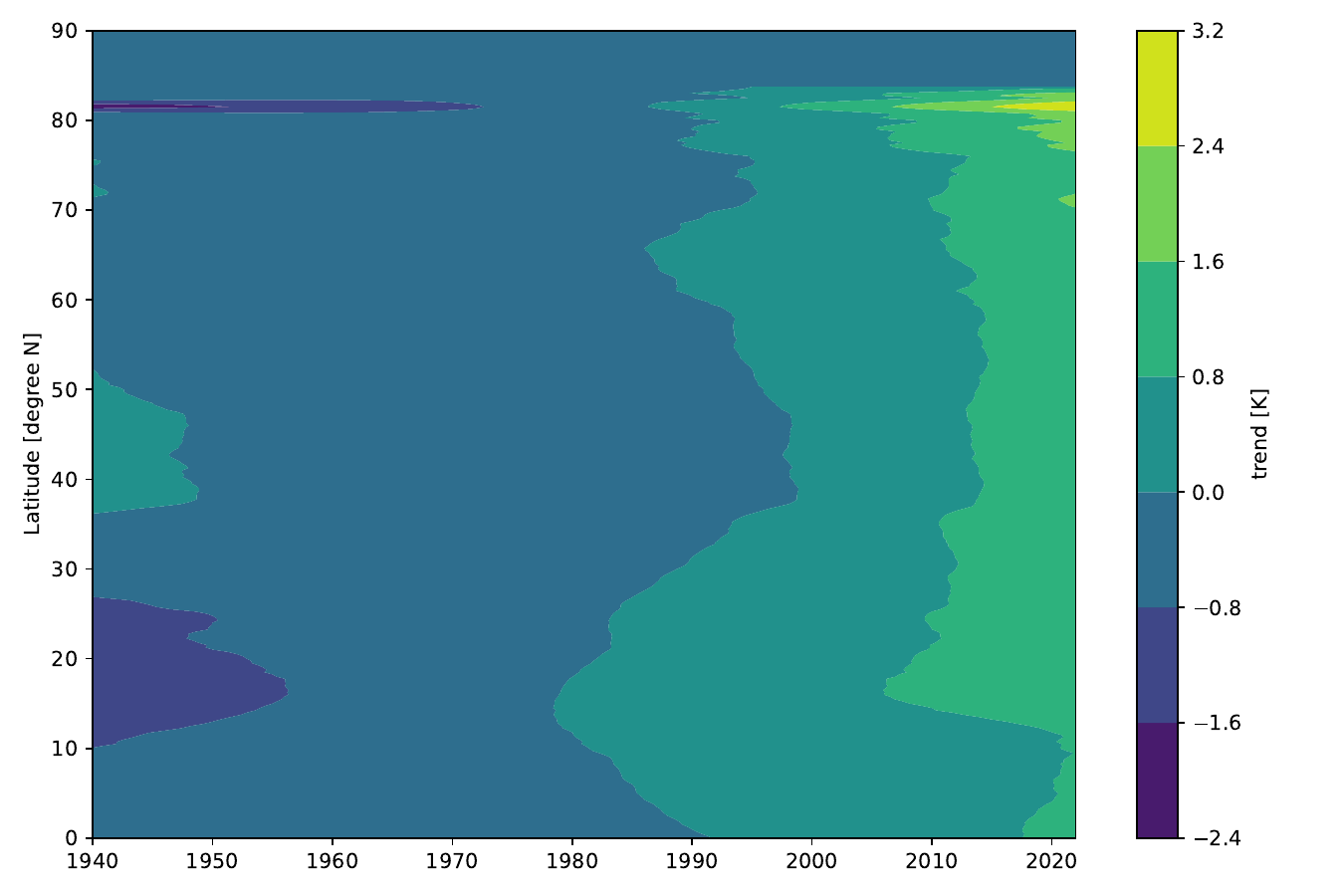}
        \caption{Contour plot of the \SI{2}{\meter} air temperature trend for ERA5 dataset as function of years and latitude. The oceans are masked.}
        \label{fig:ERA-t2m-trend}
    \end{figure}

\begin{figure}[h]
        \centering
        \includegraphics[width=0.6\textwidth]{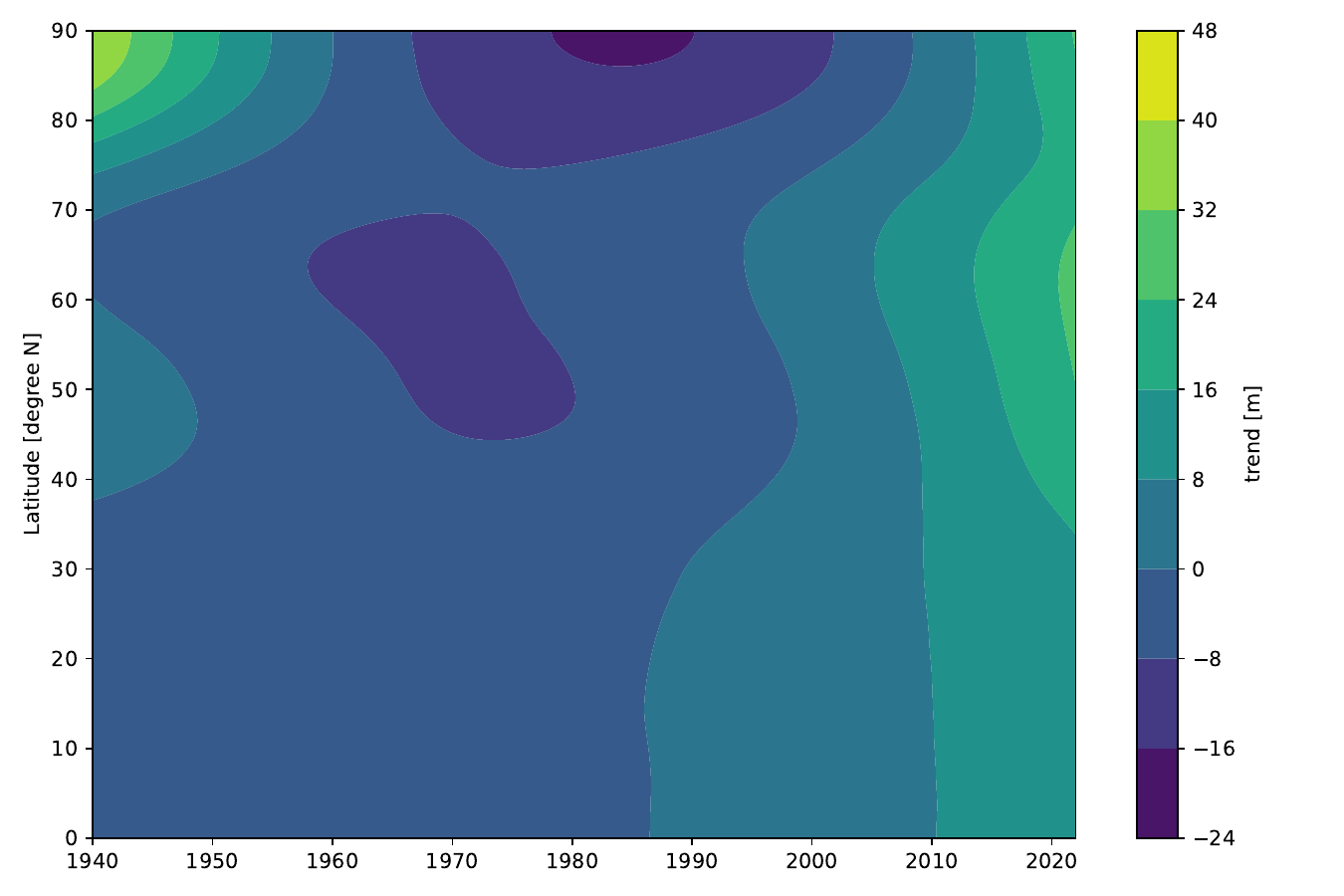}
        \caption[Contour plot of the \SI{500}{\hecto\pascal} geopotential height trend for ERA5 dataset as function of years and latitude.]{Contour plot of the \SI{500}{\hecto\pascal} geopotential height trend for ERA5 dataset as function of years and latitude. At high latitudes, the trend is non-monotonic, while it is monotonically increasing in time at lower latitudes.}
        \label{fig:ERA-geop-trend}
    \end{figure}

    In this manuscript, we present an application of the double jet index to the ERA5 reanalysis dataset \citep{hersbach_era5_2020}. The daily data are publicly available at ECMWF website (\url{https://www.ecmwf.int/en/forecasts/dataset/ecmwf-reanalysis-v5}). Given that we are interested solely in the summer period, we only downloaded the daily values of \SI{2}{\meter} air temperature ($T_\text{2m}$), \SI{500}{\hecto\pascal} geopotential height ($Z_\text{500}$), zonal wind $U$ (averaged between \SI{200}{\hecto\pascal} and \SI{350}{\hecto\pascal}) for the months of June, July and August for the Northern Hemisphere. The first step for using the ERA5 dataset consisted to a detrending process, as we want to study the response of climate in a stationary condition. We only detrend \SI{2}{\meter} air temperature ($T_\text{2m}$) and \SI{500}{\hecto\pascal} geopotential height ($Z_\text{500}$). Because we noticed a latitudinal dependency of the trend for both variables, we performed a latitudinal quadratic detrend of the seasonal zonal averages of  \SI{2}{\meter} air temperature over land only ($T_\text{2m}$) and \SI{500}{\hecto\pascal} geopotential height ($Z_\text{500}$). The contour plots of both trends are shown in \cref{fig:ERA-t2m-trend} and \cref{fig:ERA-geop-trend}. The latitudinal dependency is stronger for the  \SI{500}{\hecto\pascal} geopotential height ($Z_\text{500}$) and is present at the beginning of the dataset, suggesting a potential bias in the quality of the data before the satellite era.

    \subsection{Return time curves}
    \label{supmat:rt_curves}
    In this section we detail the computation of the return times presented in 
    \cref{fig:rt-gktl-k001}
    . We first explain how we compute the return times for the 1000 years long control simulation and then how we obtain them with the rare event algorithm. Further infomation can be found in \citep{lestang_computing_2018, ragone_computation_2018}.
    
    Given a stochastic process $ \{X(t)\}$, an observable which depends of the path $\{O[X(t)]\}$ (indicated for simplicity as $ \{O(t)\}$ from now on)  and a threshold value $a$ which separates between rare and not rare events, we can define the random variable $\tau(a,t) = \text{min}\{ \tau \geq t | O(\tau) > a \}$. Then the return time is defined as the average time to wait to see an event of magnitude higher than $a$:
        \begin{equation} \label{eq:rt_process}
            r(a) = \mathbb{E} \left[ \tau (a,t)\right]
        \end{equation}
    We can estimate the return time $r(a)$ thanks to the realization of the stochastic process, of simulation length $T_d$. This means that we have access to a finite time realization of the process and of the path-dependent observable, which we denote as $ \{O(t)\}_{0 \leq t \leq T_d}$. When we want to study high fluctuations of the stochastic process (or of any quantity which depends on it $O[X(t)]$), namely when $a$ is high, we are interested in time scales which are higher than the typical correlation time  $\tau_c$ of the process of interest, i.e. $ r(a) \gg \tau_c$. 
    Thus, the return time coincides with the time to wait, on average, between two statistically independent events both exceeding the value $a$. In \citep{lestang_computing_2018} the authors devised a methodology to correctly sample rare events based on the context presented before. Let's divide the time series of $\{O(t)\}_{0 \leq t \leq T_d}$ in $M$ blocks of duration $\Delta T_d \gg \tau_c$, such that $T_d = M\Delta T_d$. For each block, let's define the block maximum:
    \begin{equation}
        a_m = \text{max} \{O(t) | (m-1)\Delta T_d \leq t \leq m\Delta T_d\}
    \end{equation}
    and 
    \begin{equation}
        s_m = 
          \begin{cases}
            1 & \text{if } a_m \geq a \\
            0 & \text{otherwise}
        \end{cases}
    \end{equation}
        for $1\leq m \leq M$. The variable $s_m$ counts how many rare independent events are observed, i.e. $N(t) = \sum_m s(a)_{\lfloor t/\Delta T_d \rfloor}$, which are well approximate by a Poisson distribution when $ r(a) \gg \tau_c$. Then, the probability of that $a_m$ is larger than $a$ can be estimated as an empirical average of the $s_m$ over the blocks, which gives access to the return time:
    \begin{equation}
    \label{eq:rt_estimate}
        \hat{r} (a) =- \frac{\Delta T_d}{\text{ln} \left( 1 - \frac{1}{M} \sum_{m = 1}^{M} s_m (a)\right)}
    \end{equation}
    In practice, we sort the sequence $\{ a_m\}_{1 \leq m \leq M}$ in decreasing order $\{ \hat{a}_m\}_{1 \leq m \leq M}$, such that $\hat{a}_1 \geq \hat{a}_2 \geq \dots \geq \hat{a}_M$. Using \cref{eq:rt_estimate} we then associate at each threshold $\{ \hat{a}_m\}$ its respective return time $r(\{ \hat{a}_m\}) = \frac{\Delta T_d}{\text{ln} \left( 1 - \frac{m}{M}\right)}$. Finally, we can plot the couple $(r(\{ \hat{a}_m\}), \hat{a}_m)$ for $1\leq m \leq M$ as in 
    \cref{fig:rt-gktl-k001} 
    (black curve).

        For computing the return time curves for a rare events algorithm, we proceed in a very similar way. The rare event algorithm presented in this manuscript allows the sampling of rare events from an ensemble of $M$ trajectories, denoted as $ \{X_m(t)\}_{0 \leq t \leq T}$, with $1\leq m \leq M$. For each of these trajectories, we will compute $a_m = \text{max}_{0 \leq t \leq T} O[X_m(t)]$. Thus, in the particular application of return time estimation for rare events algorithm simulations, each trajectory of the algorithm plays the role of a block in the previous case. However, differently from the previous case, each maxima (trajectory) carries a weight as well. Hence, instead of the  sequences $\{ a_m\}$, we now have $\{a_m, p_m \}$ for $1\leq m \leq M$. The generalization of \cref{eq:rt_estimate} in the case of non-equiprobable blocks is:
    \begin{equation}
    \label{eq:rt_estimate_rea}
        \hat{r} (a) =- \frac{T}{\text{ln} \left( 1 - \frac{1}{M} \sum_{m = 1}^{M} p_m s_m (a)\right)}.
    \end{equation}
    In practice, to plot the return time curve, we sort in decreasing order the sequence $\{ \hat{a}_m \}$ to obtain $\{ \hat{a}_m, \hat{p}_m \}$ for $1\leq m \leq M$. We then associate for each couple $\{ \hat{a}_m, \hat{p}_m\}$ its respective return time $r(\{ \hat{a}_m\}) = \frac{T}{\text{ln} \left( 1 - \sum_{l=1}^m \hat{p}_l \right)}$, with $\sum_{l=1}^m \hat{p}_l$ being the sum of the weights for events which have an amplitude greater than $\{ \hat{a}_m \}$. This is the methodology used for retrieving the blue curve in 
    \cref{fig:rt-gktl-k001}
    . Note that, to have this curve, we also perform a second average between the 10 run of the rare events algorithm. The shadow corresponds to a standard deviation among the runs. 

    \subsection{Significance test}
    \label{supmat:ttest}
    This section describes the statistical test used for assessing the significance of the composite maps of the zonal wind $U$, \SI{2}{\meter} air temperature ($T_\text{2m}$) and \SI{500}{\hecto\pascal} geopotential height ($Z_\text{500}$). We performed a Student t-test \citep{student_probable_1908}, to test if the composite map equals the unconditioned mean. We compute the $t$ value:
    \begin{equation}
        t = \sqrt{N }\frac{\mathbb{E}[X_T | D_{ji, T} > h] - \mu}{S}
    \end{equation}
    where $N$ are the independent samples, $\mathbb{E}[X_T | D_{ji, T} > h]$ is the empirical average estimator of the composite maps , $X_T$ is the average over a season of each of the fields, $D_{ji, T}$ is the seasonal double jet, $h$ is the threshold,  $\mu$ is the unconditional mean, $S^2$ is the sample variance. We compare the $t$ value with the Student $t$ distribution value $t_{q}^N$ with $N-1$ degrees of freedom at level $q$. Thus, for a given value $q$, then we know that if $|t| \geq t_{q}^N$ then we reject the null hypothesis with probability $q$, i.e. the grid-point of the composite map is significant at a level $q$. In our case $N=10$. For our study, we want to assess significant areas in the composite maps, thus this test is applied grid point-wise.
    For the composite maps obtained with the rare events algorithm, seasonal events might still not be independent due to possible common ancestor trajectories.  When we designed the simulations, we run different sensitivity tests on the parameter $k$ to be sure that an minimum value of 10\% of the total number of trajectories in each experiment had independent ancestors at the beginning of the simulation. This can be seen in 
    fig. 11.
    Additionally, we have performed 10 experiments each starting from completely different sets of 100 initial conditions. This means that the 10 experiments are statistically independent of each other. The composite maps are obtained averaging first within each experiment, and then among the 10 experiments. The standard deviation used for the significance test is the one among the 10 estimates obtained from the 10 experiments, which are statistically independent. The statistical dependence between trajectories within  a single experiment (that in any case we have kept under control) does not influence in any way the significance analysis. 

    \subsection{Additional figures}

    \begin{figure}
        \centering
        \includegraphics[width=0.6\textwidth]{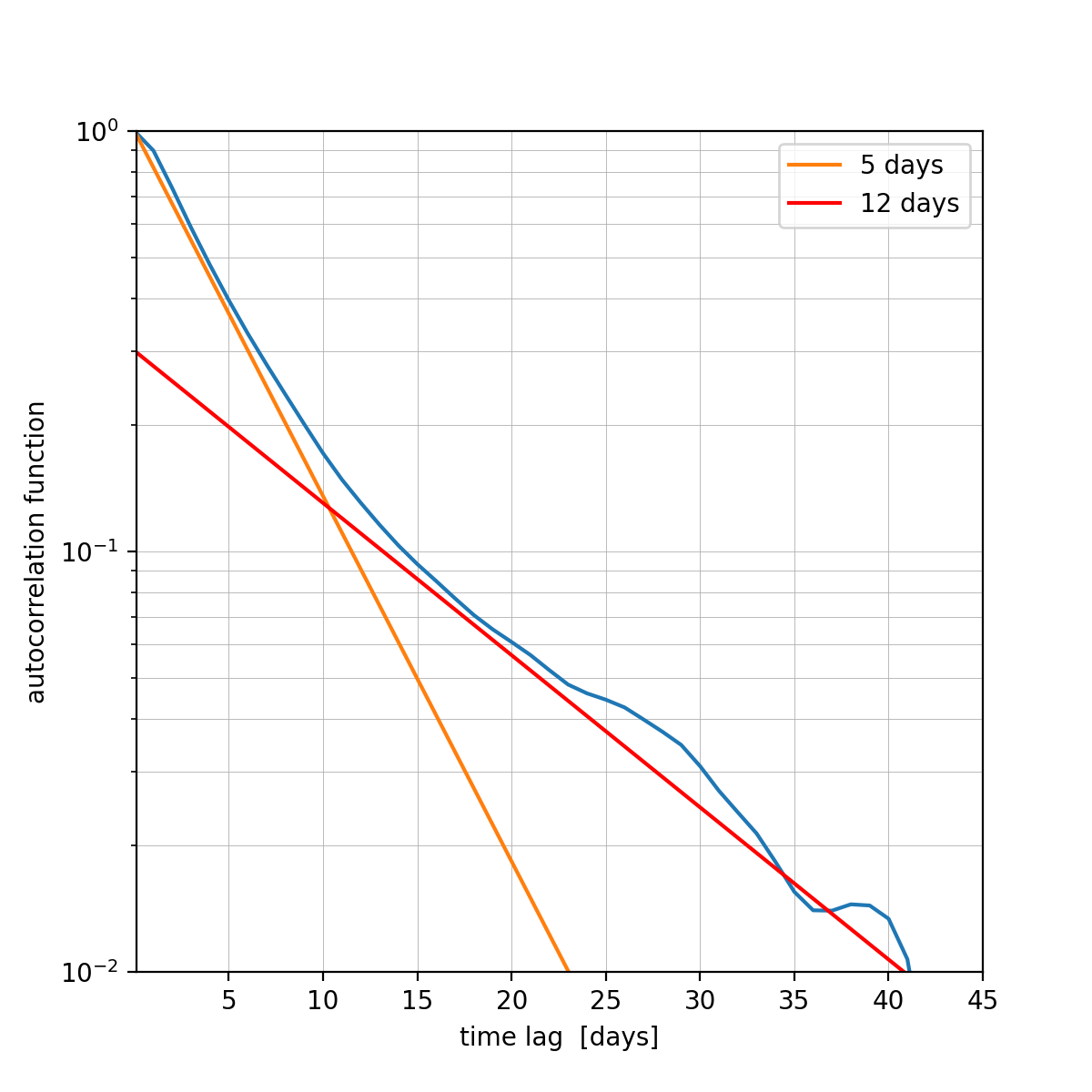}
        \caption[Autocorrelation function of the double jet index.]{Autocorrelation function of the double jet index. The orange and red lines show exponential decays on time scales of 5 and 12 days respectively.}
        \label{fig:acf-dji}
    \end{figure}

    \begin{figure}

        \centering
        \includegraphics[width=0.75\textwidth]{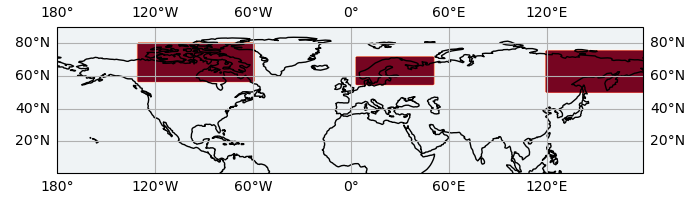}
        \caption{Regions in the Northern Hemisphere over which we computed the heatwaves: North Canada, Scandinavia and East Russia. }
        \label{fig:hw_regions}
    \end{figure}

    \begin{figure}
        \centering
        \includegraphics[width=1\textwidth]{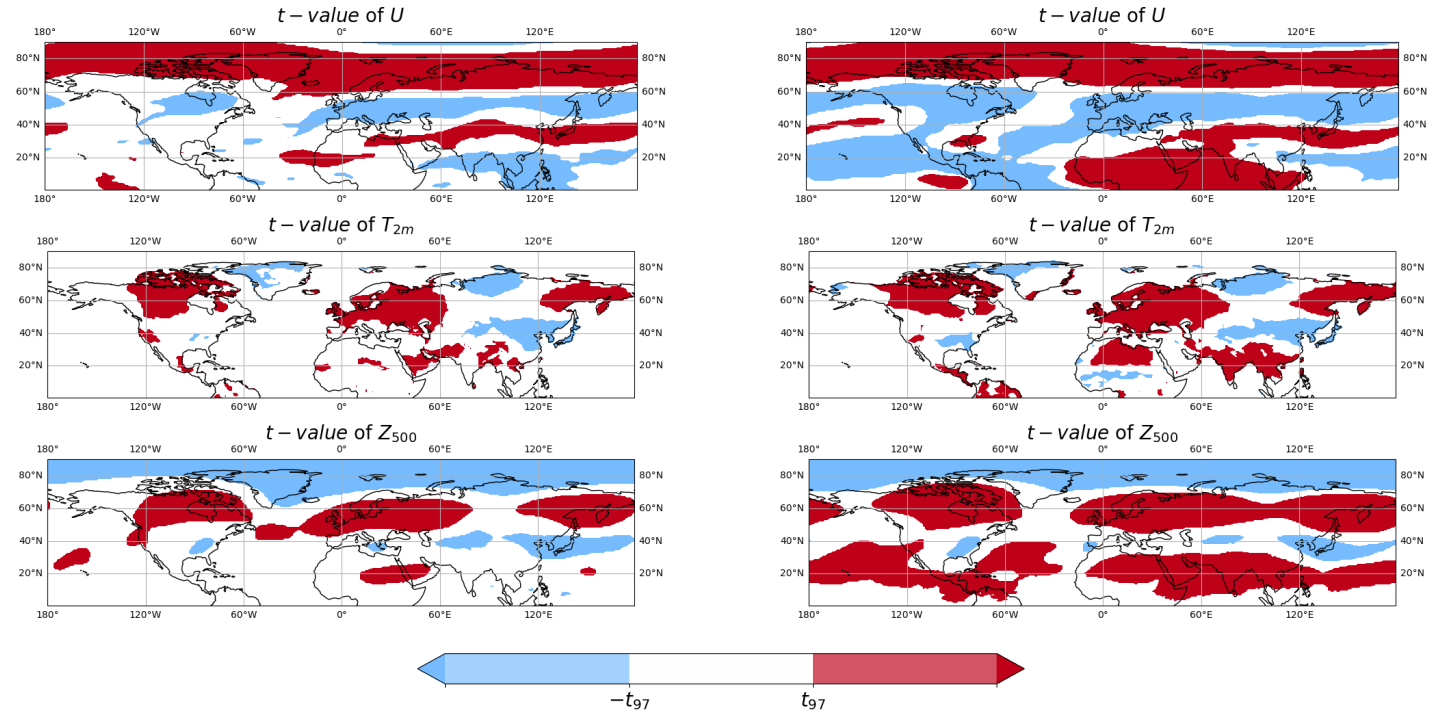}
        \caption{Significance maps for 
        \cref{fig:composite_maps_ctrl_gktl}
        .}
        \label{fig:t-values-ctrl-gktl}
    \end{figure}    

    \begin{figure}[t]
        \includegraphics[width=1\textwidth]{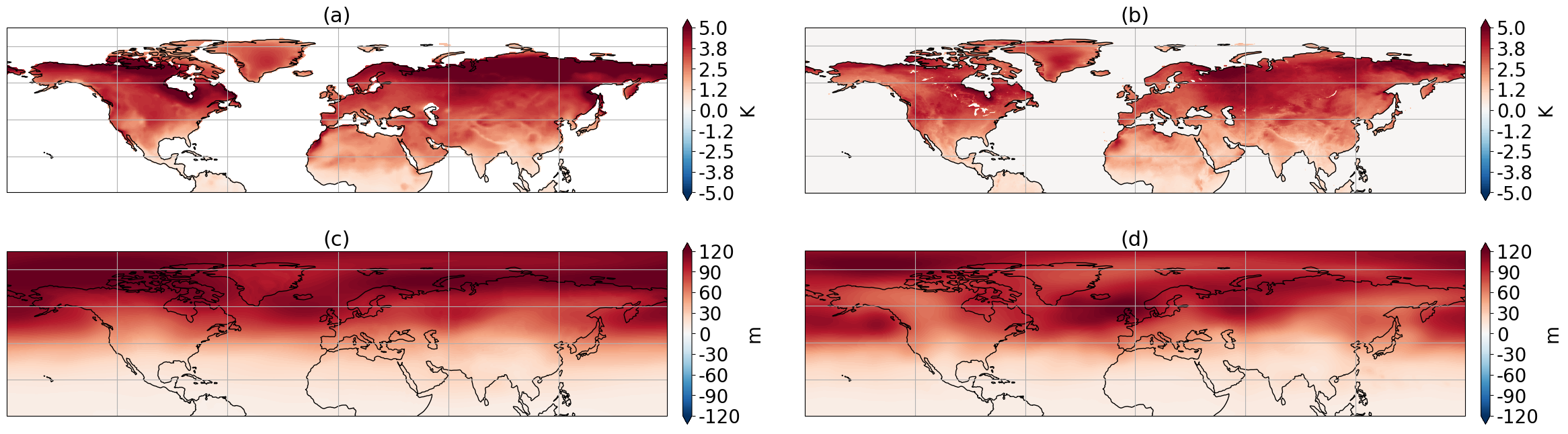}
        \caption{Standard deviation for (a) $T_{2m}$ and (b) $Z_{500}$ anomalies for 5\% most extreme double jet days for CESM1.2 control run, (c) - (d) same but for ERA5.}
        \label{fig:std_maps_cesm-era}
    \end{figure}

    \begin{figure}

        \includegraphics[width=1\textwidth]{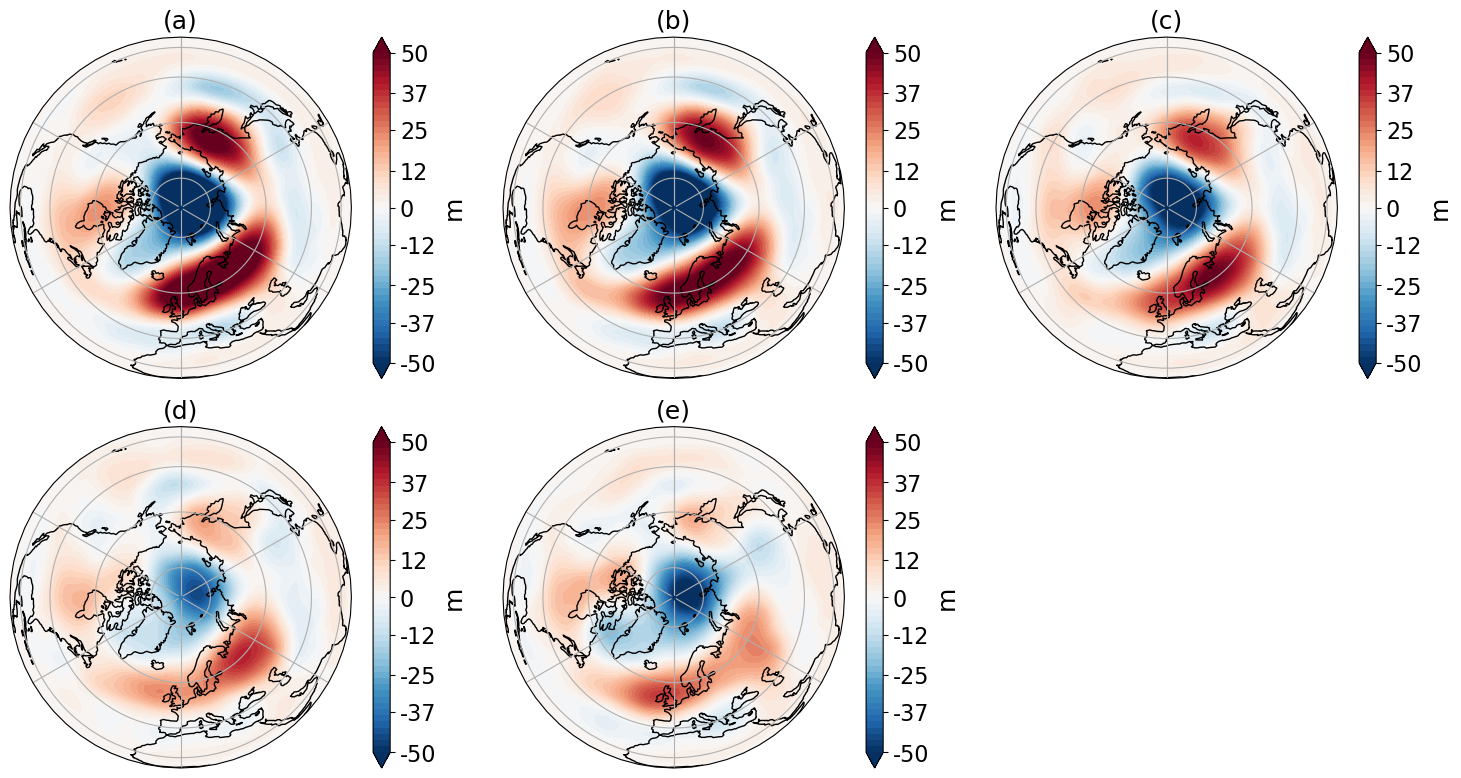}
        \caption{Composite maps for ERA5 of $Z_{500}$ anomalies for 5\% most extreme double jets, for different durations of the double jet events: (a) 1-day, (b) 7-day, (c) 14-day, (d) 30-day, (e) 90-day. No matter the duration of the double jet events, their are associated with a strong low pressure system over the polar region and strong anticyclonic anomalies over the midlatitudes.}
    \label{fig:zg500-ortho-era}
    \end{figure}

\clearpage
\bibliographystyle{ametsocV6}
\bibliography{PhD}

\end{document}